\documentclass[a4paper,aps,prdpreprintnumbers,showpacs,onecolumn,superscriptaddress,nofootinbib,amsmath,amssymb,notitlepage]{revtex4-2}

\usepackage{mathtools,leftindex,tensor,mhchem}
\usepackage{booktabs}
\usepackage{siunitx}
\usepackage{enumitem}
\usepackage{times}
\usepackage{dcolumn}
\usepackage{mathrsfs}
\usepackage{comment}
\usepackage{hyperref}
\usepackage{amsmath,accents}
\usepackage{mathtools}
\usepackage{bbm}
\usepackage{bm}
\usepackage{amsfonts}
\usepackage{amssymb}
\usepackage{mathrsfs}
\usepackage[scr=rsfs,cal=boondox]{mathalfa}
\usepackage{wasysym}
\usepackage{graphicx}
\usepackage[]{subfigure}
\usepackage{filecontents}
\usepackage[dvipsnames]{xcolor}
\usepackage{bbold}
\usepackage[scr=rsfs,cal=boondox]{mathalfa}
\usepackage{siunitx,booktabs}

\usepackage{enumitem}

\usepackage{stackengine}
\stackMath
\newcommand\tsup[2][2]{%
 \def\useanchorwidth{T}%
  \ifnum#1>1%
    \stackon[-1.3ex]{\tsup[\numexpr#1-1\relax]{#2}}{\mathchar"307E}%
  \else%
    \stackon[-1ex]{#2}{\mathchar"307E}%
  \fi%
}

\hypersetup{
    bookmarks=true,         
    pdftoolbar=true,        
    pdfmenubar=true,        
    pdffitwindow=true,     
    pdfstartview={FitH},     
    pdftitle={My title},     
    pdfauthor={author},      
    pdfsubject={Subject},    
    pdfcreator={Creator},    
    pdfproducer={Producer},  
    pdfkeywords={keyword1} 
    pdfnewwindow=true,       
    colorlinks=true ,        
    linkcolor=Blue  ,        
    citecolor=Blue,      
    filecolor=Blue,       
    urlcolor=Blue            
}

\newcommand{\arcsinh}{\operatorname{arcsinh}}
\newcommand{\arccosh}{\operatorname{arccosh}}
\newcommand{\arctanh}{\operatorname{arctanh}}

\newcommand{\ed}{\mathrm{d}}

\newcommand{\mi}{\mathrm{i}}

\newcommand{\oalpha}[1]{\accentset{\circ}{\alpha}}
\newcommand{\obf}[1]{\accentset{\circ}{\mathbf{f}}}
\newcommand{\boR}[1]{\accentset{\circ}{\mathbf{R}}}
\newcommand{\obF}[1]{\accentset{\circ}{\mathbf{F}}}
\newcommand{\obPi}[1]{\accentset{\circ}{\mathbf{\Pi}}}
\newcommand{\tL}{\bar{L}}

\newcommand{\tF}{{\bar{F}}}

\usepackage{scalerel}
\usepackage{tikz}
\usetikzlibrary{svg.path}

\definecolor{orcidlogocol}{HTML}{A6CE39}
\tikzset{
  orcidlogo/.pic={
    \fill[orcidlogocol] svg{M256,128c0,70.7-57.3,128-128,128C57.3,256,0,198.7,0,128C0,57.3,57.3,0,128,0C198.7,0,256,57.3,256,128z};
    \fill[white] svg{M86.3,186.2H70.9V79.1h15.4v48.4V186.2z}
                 svg{M108.9,79.1h41.6c39.6,0,57,28.3,57,53.6c0,27.5-21.5,53.6-56.8,53.6h-41.8V79.1z M124.3,172.4h24.5c34.9,0,42.9-26.5,42.9-39.7c0-21.5-13.7-39.7-43.7-39.7h-23.7V172.4z}
                 svg{M88.7,56.8c0,5.5-4.5,10.1-10.1,10.1c-5.6,0-10.1-4.6-10.1-10.1c0-5.6,4.5-10.1,10.1-10.1C84.2,46.7,88.7,51.3,88.7,56.8z};
  }
}

\newcommand\orcidicon[1]{\href{https://orcid.org/#1}{\mbox{\scalerel*{
\begin{tikzpicture}[yscale=-1,transform shape]
\pic{orcidlogo};
\end{tikzpicture}
}{|}}}}

\begin{document}


\title{Quasinormal modes of a static black hole in nonlinear electrodynamics
}

\author{Mohsen Fathi\orcidicon{0000-0002-1602-0722}}
\email{mohsen.fathi@ucentral.cl}
\affiliation{Centro de Investigaci\'{o}n en Ciencias del Espacio y F\'{i}sica Te\'{o}rica (CICEF), Universidad Central de Chile, La Serena 1710164, Chile}

\author{Ariel Guzmán\orcidicon{0009-0008-3844-1203}}
\email{ariel.guzman@estudiantes.uv.cl}
\affiliation{Instituto de F\'{i}sica y Astronom\'{i}a, Facultad de Ciencias, Universidad de Valpara\'{i}so,
Avenida Gran Breta\~{n}a 1111, Valpara\'{i}so, Chile}

\author{J.R. Villanueva\orcidicon{0000-0002-6726-492X}}
\email{jose.villanueva@uv.cl}
\affiliation{Instituto de F\'{i}sica y Astronom\'{i}a, Facultad de Ciencias, Universidad de Valpara\'{i}so,
Avenida Gran Breta\~{n}a 1111, Valpara\'{i}so, Chile}


\begin{abstract}

We investigate the axial electromagnetic quasinormal modes of a static, asymptotically Anti--de Sitter (AdS) black hole sourced by a nonlinear electrodynamics model of Pleba\'{n}ski type. Starting from the master equation governing axial perturbations, we impose ingoing boundary conditions at the event horizon and normalizable (Dirichlet) behavior at the AdS boundary. Following the approach of Jansen, we recast the radial equation into a linear generalized eigenvalue problem by using an ingoing Eddington--Finkelstein formulation, compactifying the radial domain, and regularizing the asymptotic coefficients. The resulting problem is solved using a Chebyshev--Lobatto pseudospectral discretization. We compute the fundamental quasinormal mode frequencies for both the purely electric ($Q_m=0$) and purely magnetic ($Q_e=0$) sectors, emphasizing the role of the nonlinearity parameter $\beta$ and the effective charge magnitude $Q$. Our results show that increasing either $\beta$ or $Q$ raises both the oscillation frequency $\omega_R$ and the damping rate $-\omega_I$, leading to faster but more rapidly decaying ringdown profiles. Nonlinear electrodynamics breaks the isospectrality between electric and magnetic configurations: magnetic modes are systematically less oscillatory and more weakly damped than their electric counterparts. For sufficiently large $\beta$ and small $Q_m$, the fundamental mode becomes purely imaginary ($\omega_R \approx 0$), in agreement with the absence of a trapping potential barrier in this regime. These findings reveal qualitative signatures of nonlinear electromagnetic effects on black hole perturbations and may have implications for high-field or high-charge astrophysical environments.

\bigskip

\noindent{\textit{keywords}}: Quasinormal modes; Nonlinear electrodynamics; Black hole perturbations; Pseudospectral method
\\

\noindent{PACS numbers}: 04.70.Bw; 04.40.Nr; 04.30.-w; 11.10.Lm   
\end{abstract}

\maketitle

\tableofcontents

\section{Introduction and Motivation}\label{sec:intro}

Although Maxwell's electromagnetic theory is one of the most beautiful and successful theories we know, its applicability is not entirely complete. At sufficiently high field strengths it exhibits well–known problems such as the divergence of the self-energy of point charges, and it becomes inadequate for describing electromagnetic phenomena in highly nonlinear media. These limitations were recognized early in the development of electrodynamics and have motivated extensive research on consistent nonlinear extensions of Maxwell’s theory.

A few years ago, the first exact solution of a rotating, charged black hole within Einstein’s general relativity (GR) coupled to a nonlinear electrodynamics (NLED) theory was reported by Garc\'{i}a~\cite{Garcia-Diaz:2021,DiazGarcia:2022jpc}. This breakthrough was the culmination of decades of progress in the field, beginning with the pioneering work of Salazar, Garc\'{i}a, and Pleba\'{n}ski ~\cite{Salazar:1987ap}, followed by Garc\'{i}a \& Ay\'{o}n-Beato~\cite{Ayon-Beato:1998hmi,Ayon-Beato:1999qin}, and many subsequent investigations~\cite{Breton:2004qa,Bronnikov:2000vy,Garcia-Salcedo:2004vjb,Breton:2005ye,Arellano:2008xw,Breton:2012yt,Breton:2015cza,Canate:2020btq,Breton:2022nqj}. A key element enabling the construction of rotating NLED black holes is the alignment of the metric null tetrad with the common eigenvectors of the electromagnetic field, which renders the stationary axisymmetric electromagnetic equations fully separable. However, this achievement comes at a price: the resulting Lagrangian density cannot be written in terms of the standard electromagnetic invariants using elementary functions.

This difficulty was recently resolved by Ay\'{o}n-Beato~\cite{ayon-beato_unveiling_2024}, who employed the NLED formulation in terms of mixed electromagnetic eigenvalues introduced in Ref.~\cite{Salazar:1987ap}. With this approach, he demonstrated that the underlying theory is fully determined and that the newly found stationary axisymmetric geometries with nonlinear charge correspond to exact self-gravitating NLED solutions. More recently, Galindo-Uriarte and Breton~\cite{Galindo-Uriarte:2024rgy}, reported the analogues of Kerr–Newman–(A)dS spacetimes in this nonlinear framework, further enriching the landscape of exact solutions in NLED. This black hole spacetime has been investigated in Refs.~\cite{AraujoFilho:2024lsi,filho_analysis_2025,Fathi:2025bvi} in connection with its astrophysical properties.

The significance of these advances lies in the fact that extremely strong electromagnetic fields naturally arise in highly magnetized compact objects such as magnetars and neutron stars, where nonlinear electrodynamic effects are expected to be non-negligible. Moreover, stationary NLED configurations provide valuable insight into the internal structure of rotating astrophysical bodies and may offer new perspectives on the resolution or avoidance of spacetime singularities in gravitational collapse.

In this work, we focus on the static sector of this family by suppressing the rotation parameter, thereby isolating the static, spherically symmetric NLED black hole that forms the seed of the full rotating solution. The resulting geometry exhibits several departures from the standard Reissner–Nordstr\"{o}m (RN) spacetime. Notably, the nonlinear parameter introduces effective corrections to the metric function that modify the causal structure, influence the number and location of horizons, and, importantly, render the spacetime asymptotically AdS. This asymptotic structure is of particular relevance in the context of Critical Gravity, where AdS black hole solutions exhibit nontrivial thermodynamical behavior and well-defined quasinormal spectra \cite{Alvarez2022,LinBravoGaeteZhang2024,BravoGaeteSantosZhang2025}.  
In AdS spacetimes, reflective boundary conditions lead to a discrete set of perturbative frequencies, making quasinormal modes (QNMs) an especially sensitive probe of the geometry. QNMs describe the characteristic response of a black hole to perturbations, and the full signal can be written as a superposition of exponentially damped oscillations, each corresponding to a distinct mode \cite{Vishveshwara1970,Press1971,Teukolsky1973,ChandrasekharDetweiler1975}.  
Every QNM is determined by a complex frequency whose real part sets the oscillation rate and whose imaginary part controls the damping timescale, while the excitation strength depends on mode-dependent coefficients \cite{Leaver1985,Leaver1986a,Leaver1986b,SunPrice1988,NollertSchmidt1992,Andersson1995,Andersson1997,NollertPrice1999,GlampedakisAndersson2001,GlampedakisAndersson2003}.  
Since these spectral quantities are uniquely determined by the mass and spin of the black hole, QNMs constitute the foundation of black hole spectroscopy \cite{Detweiler1980,Echeverria1989,Finn1992,Dreyer2004,BertiCardosoWill2006}.

In this work, we study the axial electromagnetic perturbations of this static Einstein–NLED black hole with the goal of characterizing its quasinormal spectrum. The effective potential governing these perturbations is strongly influenced by the nonlinear electromagnetic sector: the shape and height of the potential barrier, the near-horizon behavior, and the existence of trapping regions all depend on the electromagnetic configuration. Unlike linear Maxwell theory, the NLED framework breaks the electric–magnetic isospectrality, leading to distinct QNM spectra for purely electric and purely magnetic black holes. In particular, the magnetic sector may exhibit purely imaginary modes in certain regimes, corresponding to overdamped perturbations associated with the disappearance of a trapping potential.

To compute the QNMs, we adopt the ingoing Eddington–Finkelstein (IEF) pseudospectral method introduced by Jansen~\cite{jansen_2017}. This approach reformulates the radial perturbation equation into a generalized eigenvalue problem (GEVP) on a compactified radial domain, regularized at the AdS boundary. Solving the resulting GEVP via Chebyshev–Lobatto collocation yields a robust numerical scheme with exponential convergence properties, well suited for strongly nonlinear and near-extremal configurations.

Our analysis reveals characteristic imprints of nonlinear electrodynamics on the black hole ringdown. The oscillation frequency and damping rate of the fundamental mode generally increase with the nonlinear parameter and the charge magnitude. The magnetic sector consistently shows smaller real and imaginary parts of the frequency than the electric sector, and may even undergo transitions into overdamped behavior. These features constitute genuinely nonlinear effects absent in RN or any linear Maxwell-based extension.

The paper is organized as follows: 
In Sec.~\ref{sec:GRNLED} we review the Einstein–NLED action and present the field equations relevant to Pleba\'{n}ski-type theories.  
Section~\ref{sec:overview} describes the static NLED black hole geometry, analyzing its horizon structure and asymptotic properties.  
In Sec.~\ref{sec:perturbations} we derive the axial electromagnetic perturbations, obtain the master equation, and analyze the effective potential for electric and magnetic configurations.  
Section~\ref{sec:qnms} presents the pseudospectral IEF method, formulates the GEVP, and discusses the QNM spectrum.  
Finally, Sec.~\ref{sec:conclusions} summarizes our main results and discusses their physical implications for nonlinear electrodynamics and compact objects.

Throughout this work, we adopt natural units and the metric signature $(-,+,+,+)$. Whenever they appear, primes denote derivatives with respect to the radial coordinate.


\section{GR coupled to NLED}\label{sec:GRNLED}


The NLED model is defined by the action
\begin{equation}
\mathcal{S}=\frac{1}{16\pi}\int \mathrm{d}^4x\,\sqrt{-g}\,\left(R-L\right),
    \label{eq:action0}
\end{equation}
where $g$ denotes the determinant of the metric tensor $g_{\mu\nu}$, $R$ is the Ricci scalar, and $L\equiv L(F,G)$ is the Lagrangian density of the NLED field. The function $L$ depends on the electromagnetic invariant $F=\tfrac{1}{4}F_{\mu\nu}F^{\mu\nu}$, constructed from the field tensor $F_{\mu\nu}=\nabla_\mu A_\nu-\nabla_\nu A_\mu$ with $A_\mu$ the electromagnetic four-potential, and on the pseudoscalar invariant $G$ defined as
\begin{equation}
    \label{psinvg}
    G=\frac{{^{\star}F_{\mu \nu}} F^{\mu \nu}}{4}, 
    \qquad 
    ^{\star}F_{\mu \sigma} F^{\nu \sigma}= G\,\delta^{\nu}_{\mu}.
\end{equation}
The dual field tensor ${^{\star}F_{\mu\nu}}$ is given by
\begin{equation}
    \label{dualdef}
    ^{\star}F_{\mu \nu} \equiv \frac{1}{2}\sqrt{-g}\,
    \epsilon_{\mu \nu \alpha \beta} F^{\alpha \beta},
    \qquad 
    ^{\star}F^{\alpha \beta} = -\frac{1}{2\sqrt{-g}}\,
    \epsilon^{\alpha \beta \mu \nu } F_{\mu \nu},
\end{equation}
where the numerical Levi–Civita tensor $\epsilon_{\mu\nu\alpha\beta}$ corresponds to the four-Kronecker tensor \cite{Garcia-Diaz:2021}.  

The self-gravitating energy-momentum tensor associated with the NLED field is
\begin{equation}
    \label{emtens}
    -T^{\mu\nu}=L\, g^{\mu\nu} - L_F F^{\mu\sigma}F^{\nu}{}_{\sigma}
    - L_G F^{\mu\sigma} {^{\star}F^{\nu}}{}_{\sigma}
    =: L\, g^{\mu\nu} - F^{\mu\sigma}P^{\nu}{}_{\sigma},
\end{equation}
where $P_{\mu\nu}$ is the Pleba\'{n}ski tensor \cite{Salazar:1987ap}\footnote{Also known as the $p^{kl}$ field tensor in Born-Infeld theory \cite{Born:1934gh}.}, and $L_F\equiv\partial_F L$ and $L_G\equiv\partial_G L$.

Using the antisymmetric tensor $P_{\mu\nu}$ and its dual ${^{\star}P_{\mu\nu}}$,
\begin{equation}
    \label{dualp}
    ^{\star}P_{\mu \nu} = \frac{1}{2}\sqrt{-g}\,
    \epsilon_{\mu \nu \alpha \beta}\,P^{\alpha \beta},
    \qquad
    ^{\star}P^{\alpha \beta} = -\frac{1}{2\sqrt{-g}}\,
    \epsilon^{\alpha \beta \mu \nu}\,P_{\mu \nu},
\end{equation}
one introduces the invariants
\begin{equation}
    \label{invp}
    P = \frac{1}{4}P_{\mu\nu}P^{\mu\nu},
    \qquad 
    Q = \frac{1}{4}{^{\star}P_{\mu\nu}} P^{\mu\nu}.
\end{equation}

A Legendre transformation of $L(F,G)$ produces the structure function $H(P,Q)$:
\begin{equation}
    \label{legendretransf}
    L(F,G)=\frac{1}{2}F_{\mu\nu}P^{\mu\nu}-H(P,Q).
\end{equation}
Moreover, $F_{\mu\nu}$ and $P_{\mu\nu}$ are related through
\begin{eqnarray}
    \label{rel1}
    P_{\mu\nu} &=& 2\frac{\partial L}{\partial F^{\mu\nu}}
    = L_F F_{\mu\nu} + L_G\, {^{\star}F_{\mu\nu}},\\[4pt]
    \label{rel2}
    F_{\mu\nu} &=& 2\frac{\partial H}{\partial P^{\mu\nu}}
    = H_P P_{\mu\nu} + H_Q\, {^{\star}P_{\mu\nu}}.
\end{eqnarray}
The covariant field equations derived from the action \eqref{eq:action0} are
\begin{eqnarray}
    && R_{\mu\nu} - \frac{1}{2}g_{\mu\nu}R = 8\pi T_{\mu\nu},
        \label{eq:field_1}\\
    && \nabla_\nu\!\left(L_F F^{\mu\nu}\right)=0,
        \label{eq:field_2}
\end{eqnarray}
and the electromagnetic energy-momentum tensor takes the form
\begin{equation}
    \label{eq:Tmunu0}
    T_{\mu\nu} = 2\!\left(L_F F_\mu{}^{\alpha}F_{\nu\alpha}
    -\frac{1}{4}g_{\mu\nu}L\right).
\end{equation}
We now consider the static spherically symmetric line element
\begin{equation}
    \label{metrstat}
    \mathrm{d}s^2=-f(r)\,\mathrm{d}t^2+\frac{\mathrm{d}r^2}{f(r)}
    +r^2\left(\mathrm{d}\theta^2+\sin^2\theta\,\mathrm{d}\phi^2\right),
\end{equation}
expressed in the standard Schwarzschild coordinates $(t,r,\theta,\phi)$.  
A suitable ansatz for the electromagnetic four-potential of a dyonic configuration is
\begin{equation}
A_\mu = \varphi(r)\,\delta_\mu^t - Q_m \cos\theta\,\delta_\mu^\phi,
    \label{eq:A_gen}
\end{equation}
where $\varphi(r)$ is the electric scalar potential and $Q_m$ denotes the magnetic charge. This yields
\begin{eqnarray}
    && F_{tr}=-F_{rt}=-\varphi'(r),
        \qquad
        F_{\theta\phi}=-F_{\phi\theta}=-Q_m\sin\theta,\\
    && F=\frac{1}{2}\left[\frac{Q_m^2}{r^4}-\varphi'(r)^2\right].
\end{eqnarray}
Following Ref.~\cite{toshmatov_electromagnetic_2018}, we adopt the general form
\begin{equation}
f(r)=1-\frac{2m(r)}{r},
    \label{eq:f(r)_generic}
\end{equation}
and use the $rr$ and $\theta\theta$ components of Eq.~\eqref{eq:field_1} to obtain
\begin{eqnarray}
    && 2\pi r^2\!\left[4L_F(r)\,\varphi'(r)^2 - L(r)\right] + m'(r) = 0,
        \label{eq:E1}\\
    && \frac{4\pi}{r^2}\left[r^4 L(r) + 4 Q_m^2 L_F(r)\right]
       - r m''(r)= 0,
        \label{eq:E2}
\end{eqnarray}
leading to
\begin{eqnarray}
    && L(r) = 
    \frac{r^5 m''(r)\,\varphi'(r)^2 + 2Q_m^2 m'(r)}
         {4\pi r^2\!\left(Q_m^2 + r^4 \varphi'(r)^2\right)},
        \label{eq:solL}\\[4pt]
    && L_F(r) =
    \frac{r^2\!\left[r m''(r)-2m'(r)\right]}
         {16\pi\!\left(Q_m^2 + r^4 \varphi'(r)^2\right)}.
        \label{eq:solLF}
\end{eqnarray}
For $Q_m=0$ and $\varphi(r)=0$, these expressions correctly reproduce the purely electric and purely magnetic cases discussed in \cite{toshmatov_electromagnetic_2018}.  
For $m(r)=M$ ($=$ const.), one obtains $L=0$, recovering the Schwarzschild solution.

If $Q_t$ is the total charge enclosed within a sphere of radius $r$, then from Eq.~\eqref{eq:field_2} and Gauss’s law \cite{toshmatov_electromagnetic_2018},
\begin{equation}
    \label{eq:Qt}
    Q_t = r^2 L_F(r)\,\varphi'(r).
\end{equation}
Combining this with Eq.~\eqref{eq:solLF} yields
\begin{equation}
\varphi(r)=\int^r
\frac{
    r^5 m''(r) -2 r^4 m'(r)
    +\sqrt{
        r^8 \left[r m''(r)-2m'(r)\right]^2
        -1024 \pi^2 Q_m^2 Q_t^2 r^4
    }
}{32\pi Q_t r^4}\,\mathrm{d}r + c_1,
    \label{eq:phi(r)_gen}
\end{equation}
where $c_1$ is an integration constant.  
The explicit dependence of $\varphi(r)$ on $Q_m$ is a characteristic feature of NLED coupled to gravity and reflects the nontrivial interplay between electric and magnetic sectors.

If $Q_t=Q_e=\mathrm{const.}$ is a purely electric charge and $Q_m=0$, Eq.~\eqref{eq:Qt} integrates to
\begin{equation}
\varphi(r)=\frac{3m(r)-r\,m'(r)}{2Q_e}+c_1,
    \label{eq:varphi0}
\end{equation}
in agreement with \cite{toshmatov_electromagnetic_2018}. For instance, if the electric charge is linearly distributed so that $m(r)=M-Q_e^2/(2r)$, then Eq.~\eqref{eq:varphi0} gives $\varphi(r)=Q_e/(8\pi r)+c_1$, corresponding to the RN limit.

Later in this work we show that, for a nonlinear charge distribution, Eq.~\eqref{eq:Qt} must be modified accordingly, which in turn alters the resulting electric potential $\varphi(r)$.

\section{Overview of the static black hole solution in Einstein-NLED theory}
\label{sec:overview}


A static, spherically symmetric black hole endowed with NLED charge in the Einstein--NLED framework emerges by suppressing the rotation parameter in the stationary solution of Garc\'{i}a-D\'{i}az \cite{Garcia-Diaz:2021}. The resulting lapse function takes the form
\begin{equation}
    f(r) = 1 - \frac{r_s}{r} + \frac{8\pi F_0}{2} \frac{(1 - \beta r^2)^2}{r^2},
    \label{lapse}
\end{equation}
where $r_s = 2M$ corresponds to the Schwarzschild radius of a black hole with mass $M$, $F_0 = (Q_e^2 + Q_m^2)/(4\pi)$, and $\beta$ characterizes the degree of nonlinearity in the electromagnetic sector. Substituting \eqref{lapse} into the line element \eqref{metrstat} yields the exterior geometry of what we shall refer to as the NLED black hole (NLEDBH). In our unit convention, $[F_0] = \mathrm{length}^2$ and $[\beta] = \mathrm{length}^{-2}$.

\begin{figure}[h]
    \centering
    \includegraphics[width=6cm]{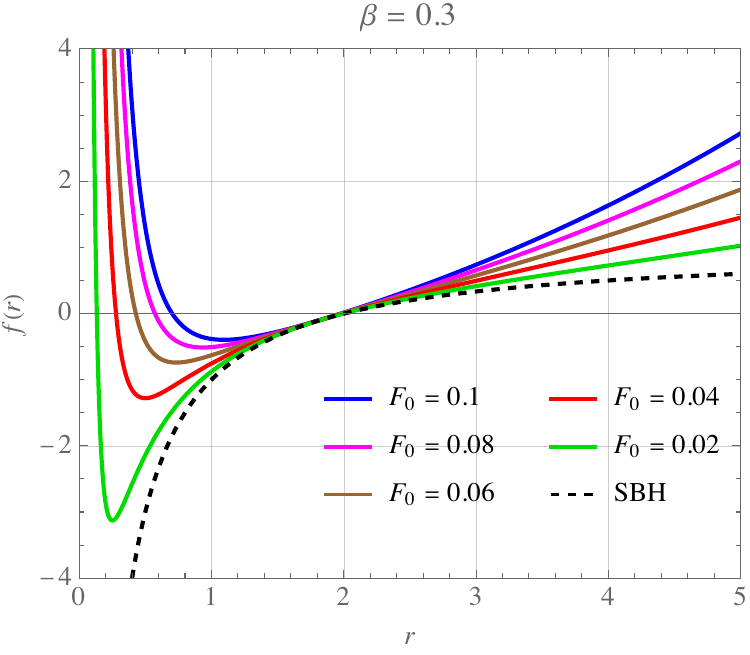} (a)\qquad
    \includegraphics[width=6cm]{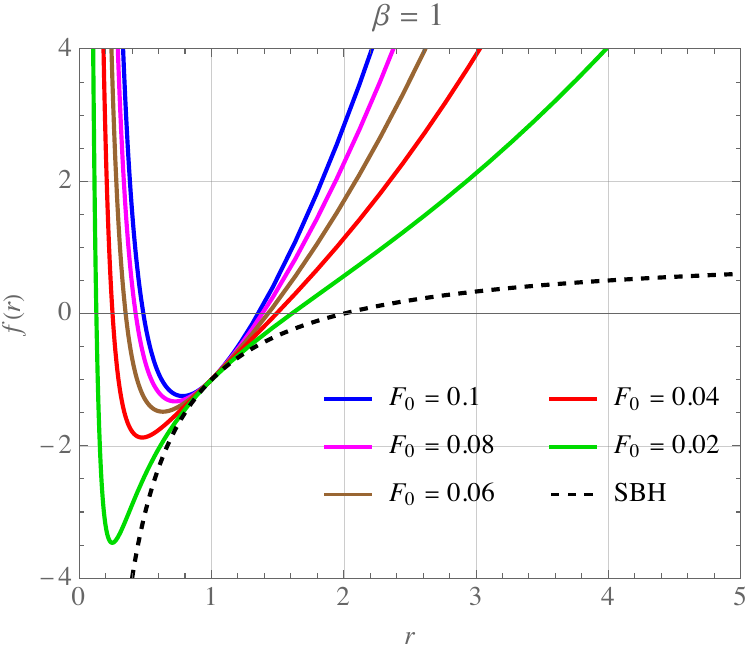} (b)
    \caption{Radial profile of the lapse function for two representative values of the nonlinearity parameter $\beta$ and several values of $F_0$. The units on both axes are normalized by the mass parameter $M$.}
    \label{fig:f(r)}
\end{figure}

It is immediate to observe that setting $\beta = 0$ recovers a linearly charged geometry, reducing the solution to the standard RN spacetime. The horizons arise from the roots of $f(r)=0$, which leads to
\begin{equation}
    r(r - r_s) = - 4\pi F_0 (1 - \beta r^2)^2.
\end{equation}
This quartic equation admits analytic solutions (see Appendix~\ref{app:A}). Since $F_0 \ge 0$, one finds that real roots occur only for $r \le r_s$. Consequently, the spacetime exhibits the structure of a Schwarzschild-Anti-de Sitter (SAdS) black hole, with an inner Cauchy horizon $r_- = r_3$ and an outer event horizon $r_+ = r_4$, while the remaining two roots form a complex conjugate pair $r_1 = r_2^*$. The lapse function can therefore be factorized as
\begin{equation}
    f(r) = \frac{4\pi \beta^2 F_0}{r^2}
    (r - r_1)(r - r_1^*)(r - r_-)(r - r_+).
    \label{eq:f(r)_new}
\end{equation}
Figure~\ref{fig:f(r)} displays the radial profile of $f(r)$ for several values of $F_0$ and two representative values of the parameter $\beta$.
The plots reveal that increasing $\beta$ systematically reduces the size of the horizons for any chosen $F_0$, although the inequality $r_+ < r_s$ always holds. Moreover, a decrease in $F_0$ causes $r_-$ to shrink and $r_+$ to expand, effectively enlarging the interior region between the horizons.

For a static metric of the form \eqref{metrstat}, the Hawking temperature associated with the event horizon is given by \cite{Misner:1973}
\begin{equation}
    T_\mathrm{H}^+ = \frac{1}{4\pi} f'(r_+),
    \label{eq:TH_0}
\end{equation}
which, upon substituting the lapse function \eqref{lapse}, yields
\begin{equation}
    T_\mathrm{H}^+ = \frac{M}{2\pi r_+^2}
    + 2 F_0 \beta^2 r_+ - \frac{2 F_0}{r_+^3}.
    \label{eq:TH_1}
\end{equation}
In the uncharged limit $F_0 = 0$, one recovers the familiar Schwarzschild temperature
\begin{equation}
    T_\mathrm{H}^+ = \frac{M}{2\pi r_s^2}.
\end{equation}
The extremal black hole (EBH) configuration corresponds to \(T_\mathrm{H}^+ = 0\), leading to
\begin{equation}
    \beta = \pm \frac{1}{r_+^2}
    \sqrt{1 - \frac{M r_+}{4\pi F_0}},
    \label{eq:beta_TH0}
\end{equation}
or equivalently,
\begin{equation}
    F_0 = \frac{M r_+}{4\pi \left(1 - \beta^2 r_+^4\right)}.
    \label{eq:F0_TH0}
\end{equation}
Equation \eqref{eq:F0_TH0} implies $F_0 > 0$ only when $\beta^2 < 1/r_+^4$. Under this condition, Eq.~\eqref{eq:beta_TH0} is automatically satisfied, indicating that extremality is permitted only within specific parameter intervals of $F_0$ and $\beta$.

\begin{figure}[h]
    \centering
    \includegraphics[width=6cm]{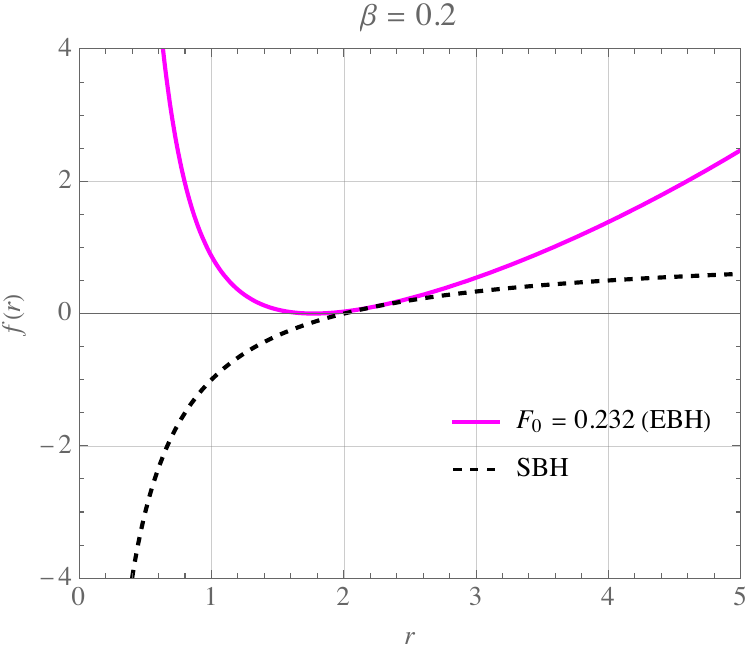} (a)\qquad
    \includegraphics[width=6cm]{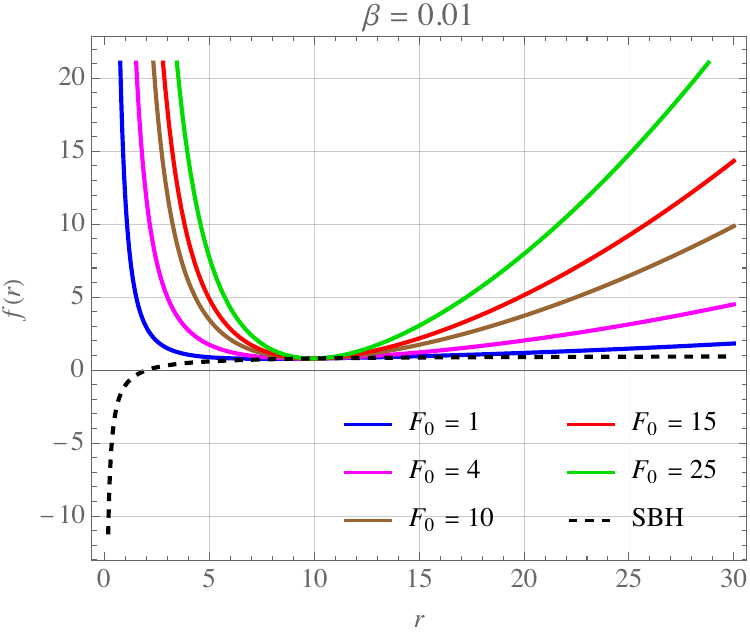} (b)
    \caption{Radial behavior of the lapse function for (a) $\beta = 0.2$, where the extremal horizon formation occurs at $r_+ = 1.771$, and (b) $\beta = 0.01$, where increasing $F_0$ leads to the disappearance of horizons and the emergence of naked singularities. Units are normalized by $M$.}
    \label{fig:f(r)_1}
\end{figure}

\begin{figure}[h]
    \centering
    \includegraphics[width=6.5cm]{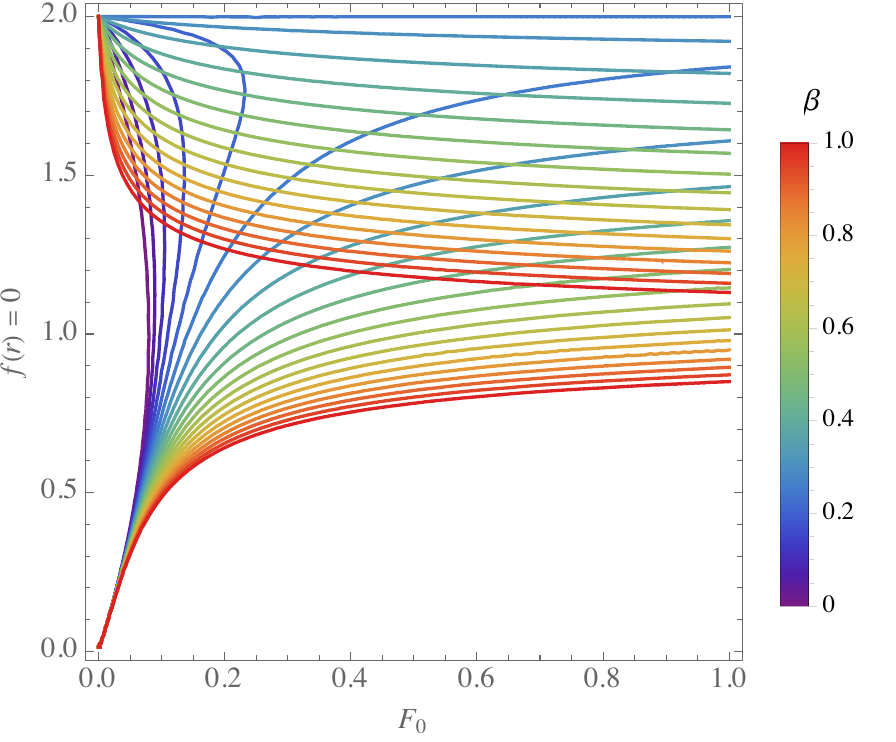} (a)\qquad
    \includegraphics[width=6.5cm]{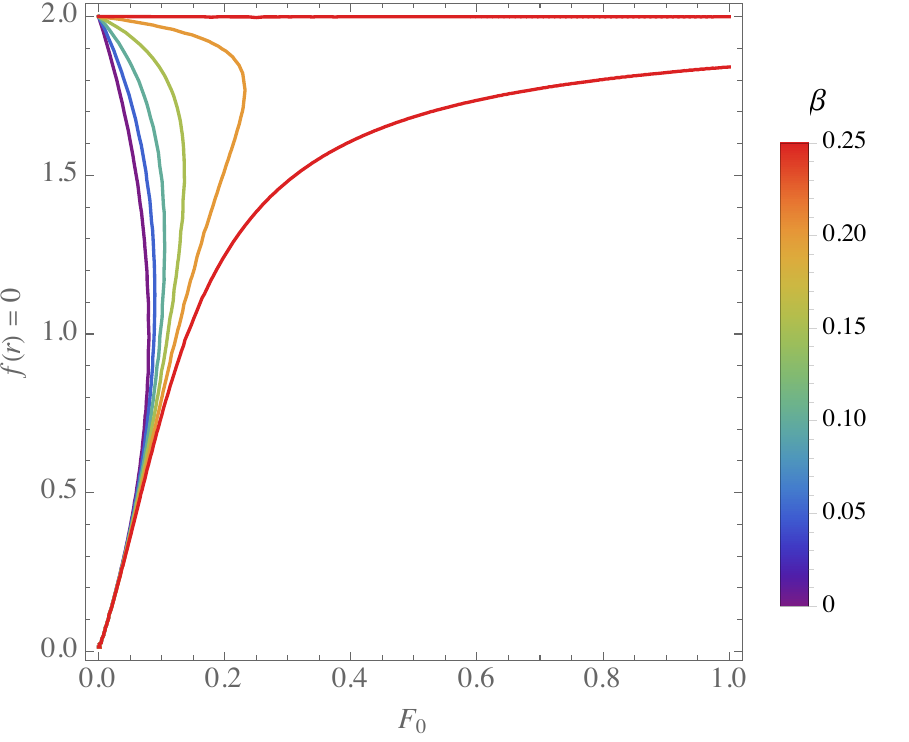} (b)
    \caption{Contours of the equation $f(r)=0$ in the $(F_0,\beta)$-plane. Panel (a) shows a broad parameter range illustrating the overall structure, while panel (b) highlights the restricted region in which extremal black holes may form.}
    \label{fig:f(r)=0}
\end{figure}

Figure~\ref{fig:f(r)_1}(a) shows that EBH configurations typically arise for moderately small values of $\beta$ and relatively large values of $F_0$. If $F_0$ becomes too large and $\beta$ too small, the extremality condition fails, the lapse function becomes strictly positive, and no horizon forms, giving rise to a naked singularity (Fig.~\ref{fig:f(r)_1}(b)).
This behavior is further illustrated in Fig.~\ref{fig:f(r)=0}, which shows the contour structure of $f(r)=0$ in the $(F_0,\beta)$-plane. Only specific parameter ranges allow the emergence of extremal horizons.
%

\section{Electromagnetic and scalar perturbations}\label{sec:perturbations}


For the NLEDBH defined through the lapse function \eqref{lapse}, the mass function takes the form
\begin{equation}
m(r) = M - \frac{2\pi F_0}{r}\left(1-\beta r^2\right)^2,
    \label{eq:m(r)_NLEDBH}
\end{equation}
so that Eq.~\eqref{eq:solLF} yields
\begin{equation}
L_F(r) = -\frac{\left(Q_e^2+Q_m^2\right)\left(1+\beta r^2\right)}{8\pi\left[Q_m^2 + r^4 \varphi'(r)^2\right]},
    \label{eq:LF_NLEDBH}
\end{equation}
where we have used the expression of $F_0$ introduced previously.  
In contrast with the RN geometry, the total charge enclosed by a sphere of radius $r$ is
\begin{equation}
Q_t \equiv Q_t(r) = 2\sqrt{\pi F_0}\left(1-\beta r^2\right)
= \sqrt{Q_e^2+Q_m^2}\left(1-\beta r^2\right),
\end{equation}
and therefore varies with $r$.  

Substituting Eq. \eqref{eq:m(r)_NLEDBH} into Eq.~\eqref{eq:phi(r)_gen}, the radial integral can be computed analytically, giving
\begin{multline}
\varphi(r) = c_2 + c_1 + \frac{\chi_0 + \sqrt{F_0}}{8\sqrt{\pi}\, r}
+ \frac{1}{8\sqrt{\pi \beta/\mu}\, \chi_0^2}\Biggl[
-2 r^2 \beta^{3/2}\sqrt{F_0/\mu}\,\chi_0^2\, 
\arctanh\!\left(\sqrt{\beta}\, r\right)\\
+\frac{\left(\sqrt{F_0}-8\sqrt{\pi}\, Q_m\right) r^2 \beta \chi_0^2}{\sqrt{\mu}}\,
\mathbf{E}\!\left(\arcsinh\!\left(\sqrt{\frac{\beta}{\mu}}\,r\right)\Bigg| \mu^2\right)
-2\sqrt{\frac{F_0}{\mu}}\, r^2 \beta \chi_0^2\,
\mathbf{F}\!\left(\arcsinh\!\left(\sqrt{\frac{\beta}{\mu}}\,r\right)\Bigg| \mu^2\right)\\
+ \frac{4 F_0 r^2 \beta \chi_0^2}{\sqrt{F_0 - 64\pi Q_m^2}}\,
\mathbf{\Pi}\!\left(-\mu;\, \arcsinh\!\left(\sqrt{\frac{\beta}{\mu}}\,r\right)\Bigg| \mu^2\right)
\Biggr],
    \label{eq:varphi_NLEDBH}
\end{multline}
where $c_2$ is an integration constant and
\begin{subequations}
\begin{align}
    &\mu = \frac{\sqrt{F_0} + 8\sqrt{\pi}\, Q_m}{\sqrt{F_0} - 8\sqrt{\pi}\, Q_m},
        \label{eq:m}\\[4pt]
    &\chi_0 \equiv \chi_0(r) =
    \sqrt{
    F_0 \left(1+\beta r^2\right)^2
    + 64\pi Q_m^2 \left(1-\beta r^2\right)^2 },
        \label{eq:chi0}
\end{align}
\end{subequations}
and $\mathbf{F}(\vartheta|\mu^2)$, $\mathbf{E}(\vartheta|\mu^2)$, and $\mathbf{\Pi}(\mathfrak{n};\vartheta|\mu^2)$ denote, respectively, the incomplete elliptic integrals of the first, second, and third kind with argument $\vartheta$, modulus $\mu^2$, and characteristic $\mathfrak{n}$ \cite{byrd_handbook_1971}.

In the linear limit $\beta\to 0$, the above expression reduces to
\begin{equation}
\varphi_l(r) = c_1 + c_2
+ \frac{\sqrt{Q_e^2 + Q_m^2}}{16\pi r}
\left(1 + \sqrt{1 + \frac{256\pi^2 Q_m^2}{Q_e^2+Q_m^2}}\right),
    \label{eq:varphi_beta_0}
\end{equation}
which further simplifies to the Coulomb profile  
\begin{equation}
\varphi(r) = c_1 + c_2 + \frac{Q_e}{8\pi r},
\end{equation}
when $Q_m=0$ and $Q_e$ is constant, reproducing the standard RN configuration.


As the black hole carries both electric and magnetic charges, we introduce the axial perturbation \cite{li_nonlinear_2015,toshmatov_electromagnetic_2018}
\begin{equation}
A_\mu = \bar{A}_\mu + \delta A_\mu,
    \label{eq:A+deltaA0}
\end{equation}
on the background gauge potential \eqref{eq:A_gen}.  
Following Ref. \cite{toshmatov_electromagnetic_2018}, the perturbation is decomposed as
\begin{equation}
\delta A_\mu = \sum_{\ell,m}
\begin{pmatrix}
0 \\
0 \\
e^{-\mi \omega t}\,\psi(r)\,\partial_\phi Y_{\ell m}(\theta,\phi)/\sin\theta \\
e^{-\mi \omega t}\,\psi(r)\,\sin\theta\,\partial_\theta Y_{\ell m}(\theta,\phi)
\end{pmatrix},
    \label{eq:deltaAmu}
\end{equation}
where $\psi(r)$ denotes the radial perturbation and $Y_{\ell m}$ are the spherical harmonics, with $\omega$ the perturbation frequency.  
The corresponding components of the field strength tensor become
\begin{subequations}
    \begin{align}
        F_{tr} &= -\varphi'(r),\\
        F_{t\theta} &= -\mi\omega\, e^{-\mi\omega t}\,\psi(r)\,\csc\theta\,\partial_\phi Y_{\ell m},\\
        F_{t\phi} &= \mi\omega\, e^{-\mi\omega t}\,\psi(r)\,\partial_\theta Y_{\ell m},\\
        F_{r\theta} &= e^{-\mi\omega t}\,\csc\theta\, \psi'(r)\,\partial_\phi Y_{\ell m},\\
        F_{r\phi} &= -e^{-\mi\omega t}\,\sin\theta\,\psi'(r)\,\partial_\theta Y_{\ell m},\\
        F_{\theta\phi} &= Q_m\sin\theta 
        - e^{-\mi\omega t}\,\psi(r)\!\left[
            \partial_\theta\!\left(\sin\theta\,\partial_\theta Y_{\ell m}\right)
            + \csc\theta\,\partial_\phi^2 Y_{\ell m}
        \right]\nonumber\\
        &= \sin\theta\Bigl[\,Q_m - \ell(\ell+1)\,e^{-\mi\omega t}\,\psi(r)\,Y_{\ell m}\Bigr].
    \end{align}
    \label{eq:Fmunu_perturbed}
\end{subequations}
To first order in the perturbation, the field strength scalar becomes
\begin{equation}
F(t,r) \approx 
\frac{1}{2}\!\left[\frac{Q_m^2}{r^4} - \varphi'(r)^2\right]
 - \frac{Q_m\,\ell(\ell+1)\,e^{-\mi\omega t}\,\psi(r)\,Y_{\ell m}}{r^4}.
    \label{eq:F_approx_0}
\end{equation}
The axial perturbations in Eq. \eqref{eq:deltaAmu} therefore influence the dynamics through the magnetic sector.  
It is convenient to decompose the field strength as  
\begin{equation}
F(t,r) = \bar{F}(r) + \delta F(t,r),
\end{equation}
where $\bar{F}(r)$ corresponds to the unperturbed contribution (the first term in Eq. \eqref{eq:F_approx_0}) and $\delta F(t,r)$ represents the perturbation (the second term in Eq. \eqref{eq:F_approx_0}).  
This modification naturally induces a perturbation in ${L}_F$, so that
\begin{equation}
L_F(t,r) = \tL_\tF(r) + \delta L_F(t,r).
    \label{eq:LF_perturbed_0}
\end{equation}
By incorporating the perturbations in Eq.~\eqref{eq:Fmunu_perturbed} into the Einstein equations \eqref{eq:field_1}, and using the mass function \eqref{eq:m(r)_NLEDBH}, we obtain
\begin{subequations}
    \begin{align}
        \tL_\tF &= \frac{(Q_e^2+Q_m^2)(1+\beta r^2)}{8\pi\left[\,Q_m^2+r^4\varphi'(r)^2\,\right]},\\[3pt]
        \delta L_F &= \frac{r^2}{8\pi}\,(Q_e^2+Q_m^2)(1+\beta r^2)
        \left(\frac{\chi_1+r^2}{r^2}\right)^2\Biggl[
        e^{-2\mi\omega t}(\chi_1+r^2)\psi'(r)^2\bigl(\partial_\theta Y_{\ell m}\bigr)^2 \nonumber\\
        &\quad -\omega^2 e^{-2\mi\omega t}r^4\csc^2\theta\,\psi(r)^2\bigl(\partial_\phi Y_{\ell m}\bigr)^2
        -\Bigl(
            \bigl[Q_m-\ell(\ell+1)e^{-\mi\omega t}\psi(r)Y_{\ell m}\bigr]^2
            + r^4\varphi'(r)^2
        \Bigr)
        \Biggr]^{-1},
    \end{align}
    \label{eq:tLF_deltaLF}
\end{subequations}
where $\chi_1\equiv\chi_1(r)=(Q_e^2+Q_m^2)(1-\beta r^2)^2-2Mr$.  
These expressions represent the unperturbed contribution to $L_F$, corresponding to Eq.~\eqref{eq:LF_NLEDBH}, and its perturbation.  
It is also straightforward to verify that $\delta L_F=\tL_{\tF\tF}\,\delta F$ \cite{toshmatov_electromagnetic_2018}, where $\tL_{\tF\tF}\equiv\partial_{\tF}\tL_{\tF}$.

From the covariant equation of motion \eqref{eq:field_2}, the axial electromagnetic perturbation satisfies the wave equation \cite{li_nonlinear_2015}
\begin{equation}
\frac{\ed^2\psi}{\ed r_*^2}
+ \bigl[\omega^2 - V_{\mathrm{em}}(r)\bigr]\psi(r)=0,
    \label{eq:psi_eq_0}
\end{equation}
expressed in terms of the tortoise coordinate, $\ed r_*=\ed r/f(r)$.  
Using the lapse function \eqref{eq:f(r)_new}, the tortoise coordinate becomes
\begin{multline}
r_* = \frac{1}{\beta^2(Q_e^2+Q_m^2)}\Biggl[
\frac{r_1^2\,\ln(r-r_1)}{(r_1-r_1^*)(r_1-r_-)(r_1-r_+)}
-\frac{{r_1^*}^2\,\ln(r-r_1^*)}{(r_1-r_1^*)(r_1^*-r_-)(r_1^*-r_+)}\\
-\frac{r_-^2\,\ln(r-r_-)}{(r_1-r_-)(r_1^*-r_-)(r_+-r_-)}
-\frac{r_+^2\,\ln(r-r_+)}{(r_1-r_+)(r_1^*-r_+)(r_+-r_-)}
\Biggr].
    \label{eq:r*}
\end{multline}
In Eq.~\eqref{eq:psi_eq_0}, the electromagnetic effective potential is defined as \cite{toshmatov_electromagnetic_2018}
\begin{equation}
V_\mathrm{em}(r) = f(r)\Biggl[
\frac{\ell(\ell+1)}{r^2}\left(1+\frac{4Q_m^2\tL_{\tF\tF}(r)}{r^4}\right)
-\frac{f(r)\tL_\tF'(r)^2 - 2\tL_\tF(r)\bigl(f(r)\tL_\tF'(r)\bigr)'}{\tL_\tF(r)^2}
\Biggr],
    \label{eq:V(r)_LF}
\end{equation}
which incorporates the contributions from both electric and magnetic charges as well as the NLED corrections.  

For comparison, in the linear Maxwell case, or for uncharged black holes, the effective potential takes the well-known form \cite{nollert_quasinormal_1999,konoplya_quasinormal_2011}
\begin{equation}
V(r)=f(r)\left[\frac{\ell(\ell+1)}{r^2}+(1-s)\frac{f'(r)}{r}\right],
    \label{eq:V(r)_1}
\end{equation}
which applies to scalar, electromagnetic, and axial perturbations, with $s$ denoting the field spin.  
In particular, $s=0$ and $s=1$ correspond to scalar and electromagnetic perturbations, respectively, with the usual requirement $\ell\geq s$.

In the absence of nonlinear contributions, i.e. for constant $\tL_\tF$, the potential \eqref{eq:V(r)_LF} appropriately reduces to
\begin{equation}
V_\mathrm{em}(r)=\frac{\ell(\ell+1)}{r^2}f(r),
    \label{eq:V(r)_old}
\end{equation}
in agreement with the $s=1$ limit of Eq.~\eqref{eq:V(r)_1}.  
This expression applies to black hole solutions without nonlinear or magnetic corrections.

\begin{figure}[t]
    \centering
    \includegraphics[width=6.5cm]{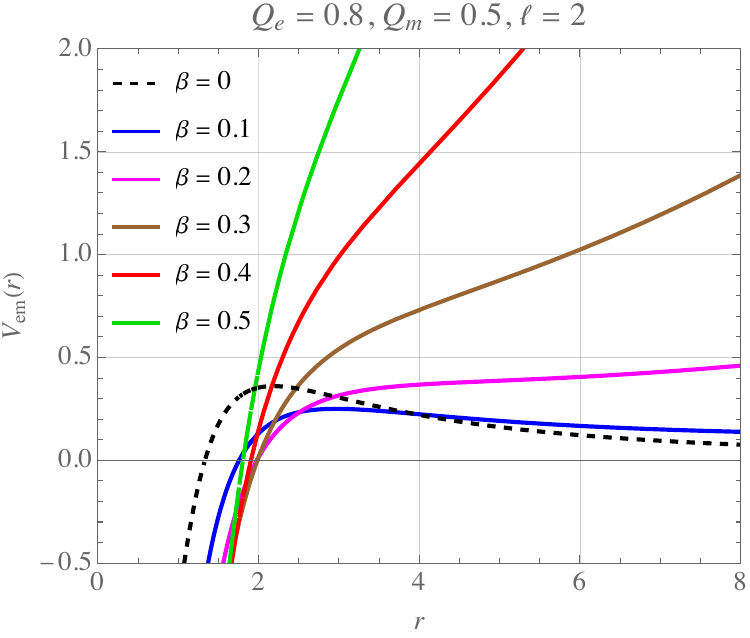} (a)\qquad
    \includegraphics[width=6.5cm]{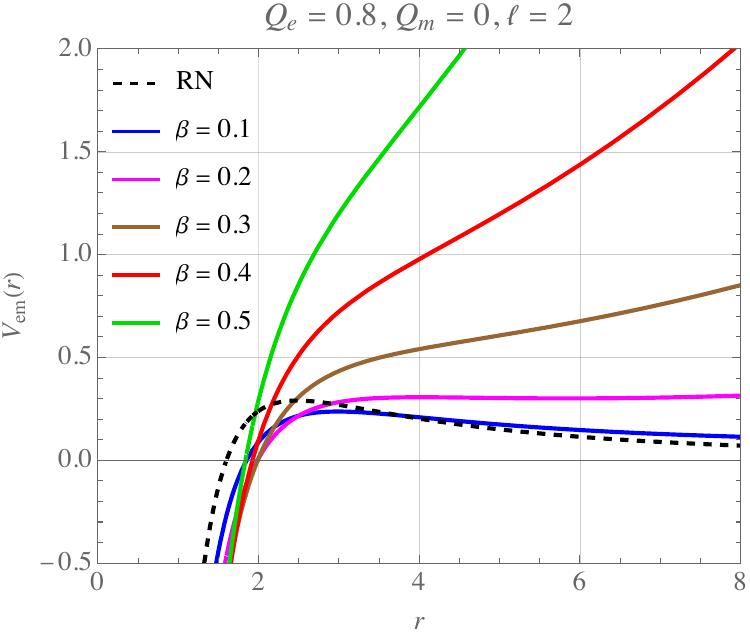} (b)
    \includegraphics[width=6.5cm]{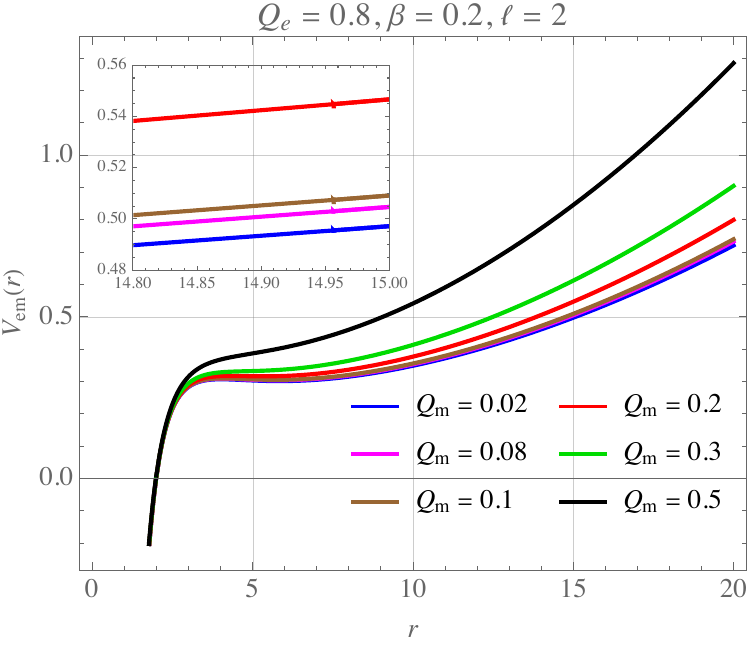} (c)\qquad
    \includegraphics[width=6.5cm]{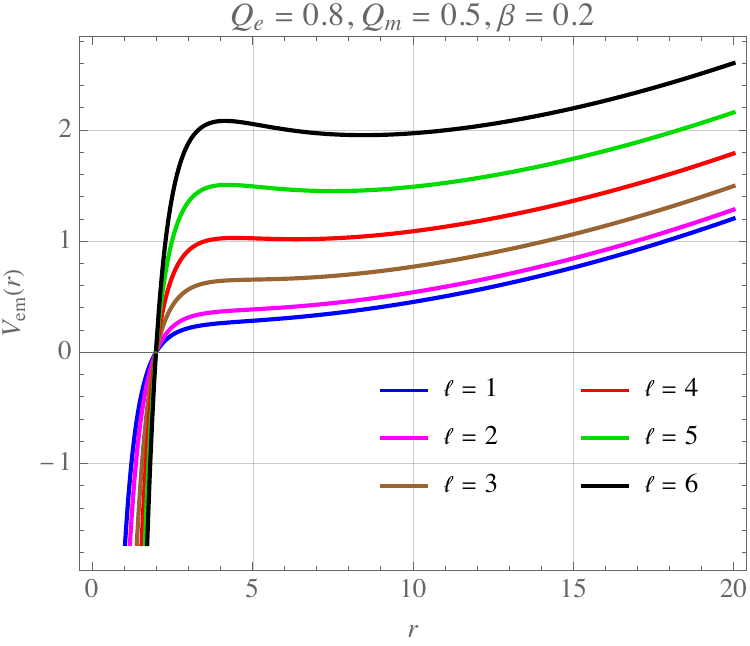} (d)
    \includegraphics[width=6.5cm]{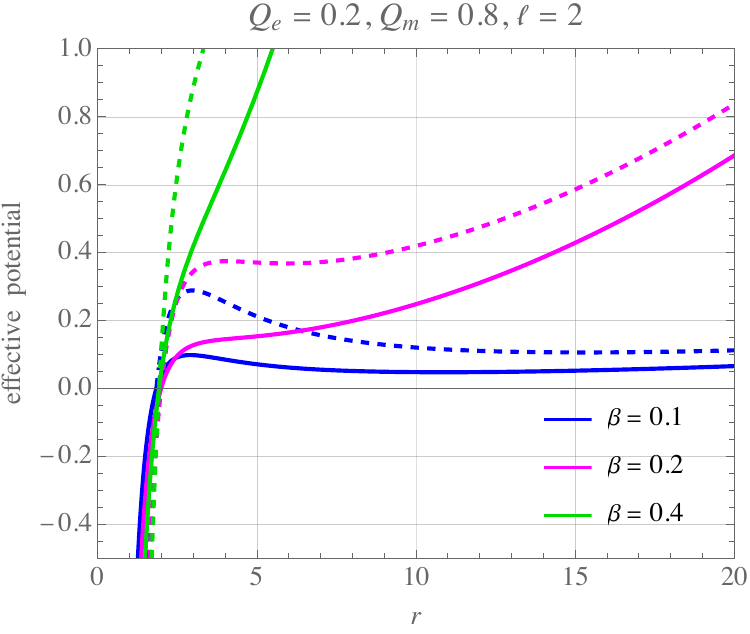} (e)\qquad
    \includegraphics[width=6.5cm]{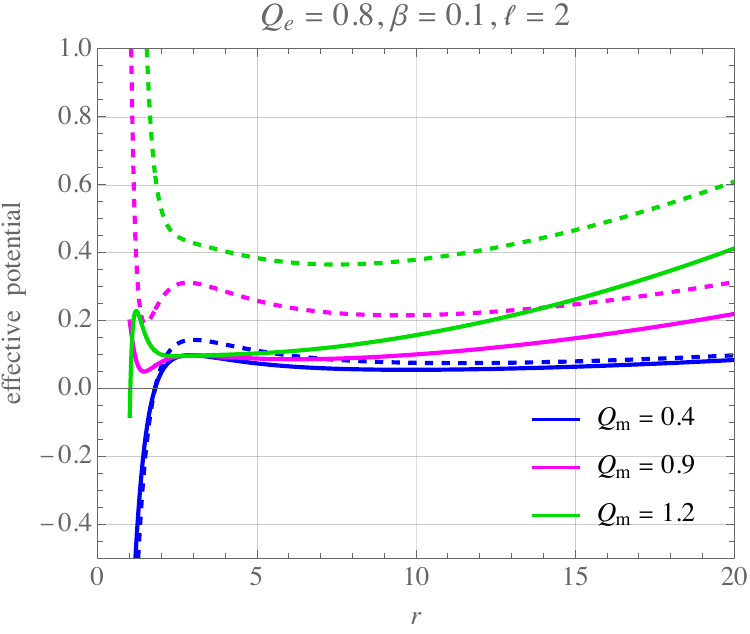} (f)
    \caption{Panels (a-d): radial profiles of the electromagnetic potential for (a) fixed electric/magnetic charges with varying nonlinearity; (b) purely electric black holes with varying nonlinearity; (c) varying magnetic charge; and (d) fixed nonlinearity with different multipole index $\ell$.  
    Panels (e-f): comparison between the electromagnetic potential $V_{\mathrm{em}}(r)$ and the scalar potential $V(r)$ for (e) fixed charges with varying nonlinearity and (f) fixed electric charge and nonlinearity with varying magnetic charge.  
    Solid curves correspond to scalar perturbations and dashed curves to electromagnetic perturbations, with $\ell=2$ in all cases.  
    The mass $M$ is used as the unit of length.}
    \label{fig:Vem}
\end{figure}

Figure~\ref{fig:Vem} displays several examples of the potential $V_{\mathrm{em}}(r)$ under different configurations.
As shown in the plots, increasing the nonlinearity parameter $\beta$ significantly enhances the height of the potential barrier and produces a non-decaying asymptotic behavior.  
For small $\beta$, the potential remains damped and approaches the RN form, particularly for purely electric configurations.  
At fixed $\beta$, an increase in the magnetic charge raises the peak of the potential, and similarly, increasing the multipole number $\ell$ amplifies the barrier.  

Finally, by comparing electromagnetic and scalar perturbations, one sees that stronger nonlinearity leads to a more pronounced difference between their potential barriers, especially in the asymptotic region.



\section{QNM{s} of the black hole}
\label{sec:qnms}

\begin{figure}[t]
    \centering
    \includegraphics[width=7.0cm]{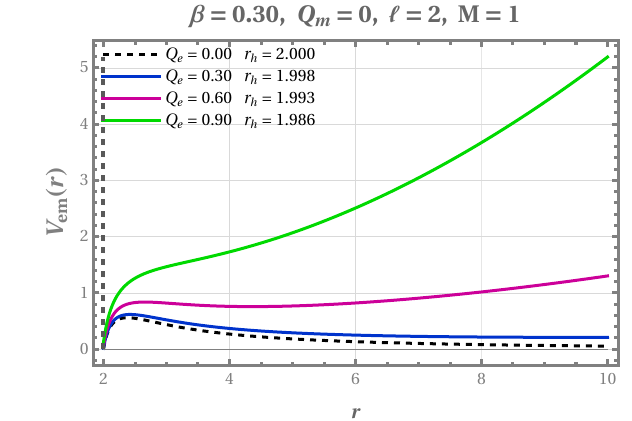} (a)\qquad
     \includegraphics[width=7.0cm]{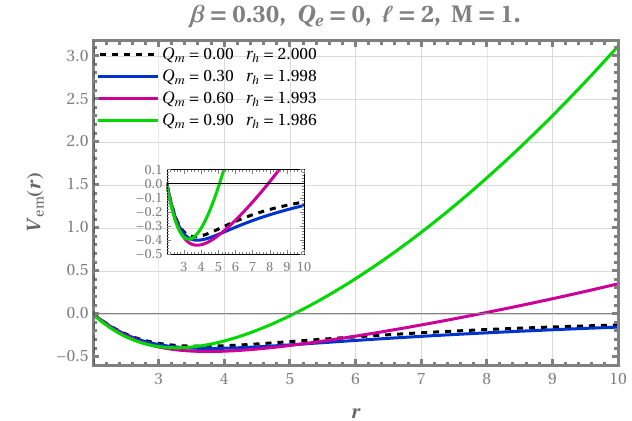} (b)
      \includegraphics[width=7.0cm]{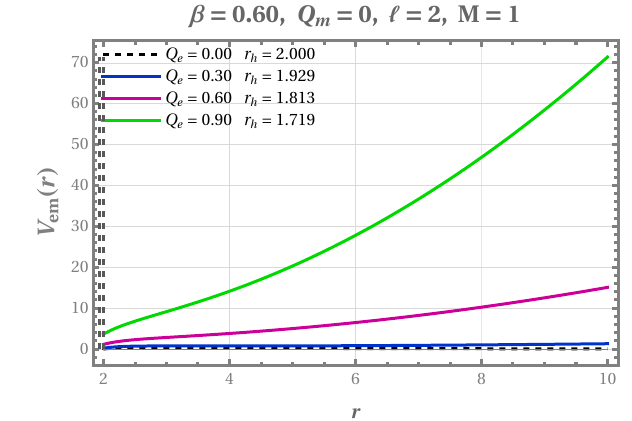} (c)\qquad
       \includegraphics[width=7.0cm]{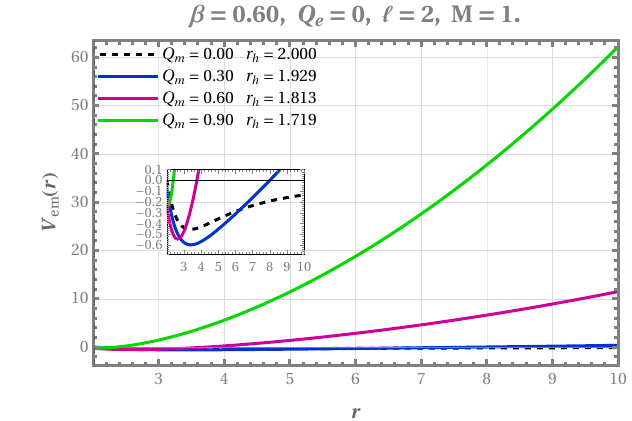} (d)
       \includegraphics[width=7.0cm]{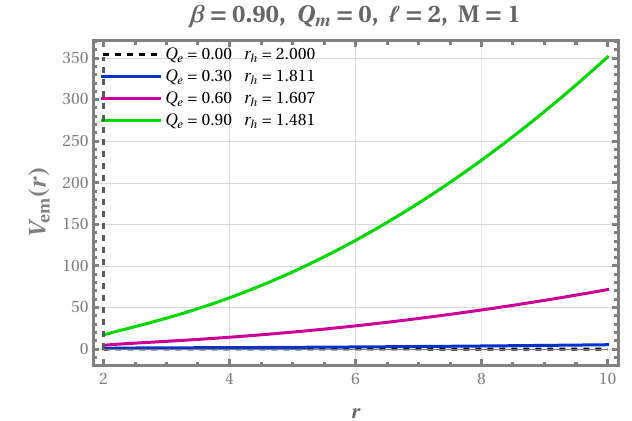} (e)\qquad
       \includegraphics[width=7.0cm]{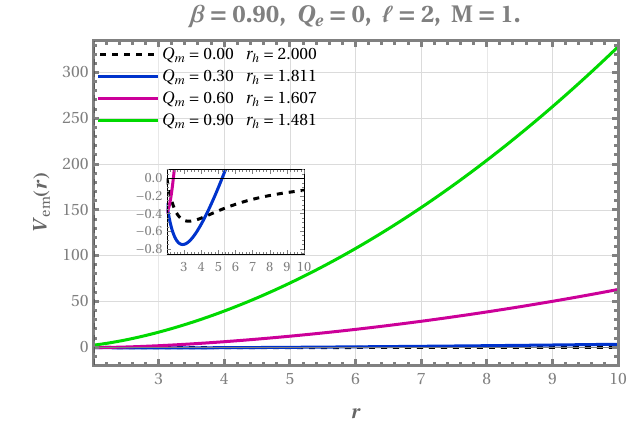} (f)
       
    \caption{Radial profiles of the effective potential for purely electric and purely magnetic configurations. Each row corresponds to a fixed nonlinearity parameter, with $\beta=0.30$ (a,b), $0.60$ (c,d), and $0.90$ (e,f). The left column shows purely electric cases ($Q_m=0$), while the right column presents purely magnetic ones ($Q_e=0$). Within each panel, the curves denote $Q\in\{0.3,0.6,0.9\}$ for $M=1$. The vertical marker indicates the horizon radius $r_+$, and the horizontal line marks $V_{\mathrm{em}}=0$. The spherical-harmonic index is fixed to $\ell=2$, and all length scales are normalized by the black hole mass $M$. 
    In the electric sector (a,c,e), the potential develops a horizon-centered barrier followed by the characteristic NLED-induced growth at large $r$. In the magnetic sector (b,d,f), the NLED angular contribution suppresses the near-horizon barrier and may produce a shallow well; for $\beta=0.90$ and $Q_m=0.30$, the potential becomes monotonic, with neither barrier nor well.}
    \label{fig:Vem_1}
\end{figure}

In this section, we compute the QNMs associated with the static NLEDBH. Starting from the master equation for axial electromagnetic perturbations, we impose the appropriate physical boundary conditions and recast the problem into a \textit{linear generalized eigenvalue formulation}, following the pseudospectral framework developed by Jansen~\cite{jansen_2017}. We then outline the full numerical procedure, which includes the compactification of the radial coordinate, the regularization of both the near-horizon and asymptotic behavior, the construction of the collocation grid, the identification of physical modes through spectral filtering, and a rigorous assessment of numerical convergence. Finally, we compute the fundamental QNM spectra for the \textit{purely electric} and \textit{purely magnetic} sectors, examining their dependence on the nonlinear parameter $\beta$ and on the effective electromagnetic charge. Note that, in Fig.~\ref{fig:Vem_1}, representative profiles of the effective potential $V_{\mathrm{em}}(r)$ in Eq.~\eqref{eq:V(r)_LF} are displayed for both purely electric and purely magnetic configurations.
%

\subsection{Master equations 
}
\label{subsec:master_potential}

To analyze axial electromagnetic perturbations, we recall that the radial master field $\psi(r)$ satisfies the Regge--Wheeler--type Schr\"{o}dinger equation \eqref{eq:psi_eq_0}. Throughout this section, the functions $L_F$ and $L_{FF}$ are evaluated on the static NLEDBH background.\footnote{For axial perturbations in nonlinear electrodynamics, the term proportional to $Q_m^2 L_{FF}$ contributes solely to the purely magnetic sector, whereas the structural NLED term $\mathcal{C}_{\mathrm{NLED}} = \bigl[f L_F'^2 - 2L_F(fL_F')'\bigr]/L_F^2$ appears in both purely electric and purely magnetic configurations; see Refs.~\cite{toshmatov_electromagnetic_2018,toshmatov_polar_2018}.}

It should, however, be borne in mind that the following items must be taken into consideration:

\begin{itemize}

\item \textbf{\textit{Domain of validity}.} 
We require that $L_F(r)\neq 0$ and that all coefficients in Eq.~\eqref{eq:psi_eq_0} remain finite for $r\in[r_+,\infty)$, ensuring that the master equation is hyperbolic and free of nonphysical singularities~\cite{Schellstede:2016}. This condition motivates the parameter domain adopted later (Subsec.~\ref{sec:pipeline}): any configuration in which $L_F$ or $L_{FF}$ vanishes or diverges outside the horizon is excluded from the QNM analysis.

\item \textbf{\textit{Linear (Maxwell) limit and sector split}.}
In the Maxwell limit $L_F\to 1$ and $L_{FF}\to 0$, the effective potential reduces to  
that in Eq. \eqref{eq:V(r)_old},  
reproducing the standard axial electromagnetic potential in RN or Schwarzschild backgrounds; see Ref.~\cite{konoplya_quasinormal_2011}. In NLED, however, the electric and magnetic sectors cease to be isospectral. We explicitly implement this sectoral split in Subsec.~\ref{sec:sectors}.

\item \textbf{\textit{Scope}.}  
The boundary conditions (ingoing at $r=r_+$ and normalizable at the asymptotic AdS boundary), together with the IEF reformulation, are presented in Subsecs.~\ref{sec:boundary_conditions}--\ref{sec:ef_compactification}. There, we cast the perturbation equation into a linear GEVP suitable for pseudospectral discretization, following the framework of Refs.~\cite{jansen_2017,konoplya_quasinormal_2011,toshmatov_polar_2018,horowitz_hubeny_2000}.

\end{itemize}
%

\subsection{Boundary conditions and IEF reformulation}
\label{sec:boundary_conditions}

We impose an \emph{ingoing-wave} boundary condition at the horizon $r=r_+$ and a \emph{normalizable} (Dirichlet-type) condition at the asymptotic AdS boundary. To guarantee regularity at the horizon, we employ the IEF chart, defined through the advanced time coordinate
\begin{equation}
v = t + r_*,
\end{equation}
which renders the metric manifestly regular across $r=r_+$ and ensures that the ansatz $\psi(r)\,e^{-i\omega v}$ automatically encodes the physically required ingoing behavior. At spatial infinity, the AdS asymptotics select the decaying (normalizable) branch of the radial solution, thereby producing a discrete quasinormal-mode spectrum, in accordance with the standard AdS prescription~\cite{horowitz_hubeny_2000}.

\subsection{IEF reformulation and compactification}
\label{sec:ef_compactification}

We introduce the change of variable
\begin{equation}
r(z) \doteq \frac{r_+}{1 - z}, \qquad z \in [0,1),
\label{eq:r(z)}
\end{equation}
such that $z=0$ corresponds to the horizon $r=r_+$, while the limit $z \to 1^-$ maps to the asymptotic region $r \to +\infty$.
In this formulation, we express the scalar field in IEF coordinates as
\begin{equation}
\Psi(z) = \left(1 - z\right)\Phi(z),
\label{eq:Psi}
\end{equation}
so that $\Psi$ satisfies a Dirichlet condition at $z=1$ while $\Phi$ remains finite for $0 \le z < 1$. With these definitions, the axial master equation \eqref{eq:psi_eq_0} is cast into a GEVP, following Jansen~\cite{jansen_2017},
\begin{equation}
\mathcal{A}[\Phi] + \mathrm{i}\Lambda\,\mathcal{B}[\Phi] = 0,
\label{eq:genEVP}
\end{equation}
where $\Lambda := {\omega}/{f'(r_+)}$ and $\mathcal{A}$, $\mathcal{B}$ are second- and first-order differential operators acting on $\Phi(z)$:
\begin{subequations}
\begin{align}
& \mathcal{A}[\Phi] = a_2(z)\,\Phi'' + a_1(z)\,\Phi' + a_0(z)\,\Phi,\label{eq:APhi}\\
& \mathcal{B}[\Phi] = b_1(z)\,\Phi' + b_0(z)\,\Phi, \label{eq:BPhi}
\end{align}
\end{subequations}
with all $r$-dependent quantities rewritten using the definition~\eqref{eq:r(z)}. The explicit coefficients are
\begin{subequations}
\begin{align}
a_0(z) &= \frac{2 f(r)}{r_+^2}\left(1-z\right)^3
        - \frac{f'(r)}{r_+}\left(1-z\right)^2
        - S(z)\left(1-z\right),\\[4pt]
a_1(z) &= -\frac{4 f(r)}{r_+^2}\left(1-z\right)^4
         + \frac{f'(r)}{r_+}\left(1-z\right)^3,\\[4pt]
a_2(z) &= \frac{f(r)}{r_+^2}\left(1-z\right)^5,\\[4pt]
b_0(z) &= \frac{2 f'(r_+)}{r_+}\left(1-z\right)^2,\\[4pt]
b_1(z) &= -\frac{2 f'(r_+)}{r_+}\left(1-z\right)^3,
\end{align}
\end{subequations}
where
\begin{equation}
S(z) = \frac{\ell(\ell+1)}{r(z)^2} - \mathcal{C}_{\mathrm{NLED}}(z),
\end{equation}
and
\begin{equation}
\mathcal{C}_{\mathrm{NLED}}(z) =
\frac{f L_F'^2 - 2 L_F (f L_F')'}{L_F^2}\Bigg|_{r=r(z)}.
\label{eq:CNLED}
\end{equation}
The rescaled eigenvalue $\Lambda$ significantly improves the numerical conditioning of the discretized GEVP, particularly in the near-extremal regime~\cite{jansen_2017}.

\subsection{Asymptotic regularization as $z \to 1^-$}
\label{sec:regularization}

Because the metric function grows as $f(r)\sim \mathrm{const}\times r^{2}$ for $r\to\infty$ in the present AdS background, the raw coefficients of the GEVP \eqref{eq:genEVP} contain terms that either vanish or diverge as $z\to 1$. Following the IEF--pseudospectral strategy of Ref.~\cite{jansen_2017}, we multiply the entire Eq. \eqref{eq:genEVP} by the weight
\begin{equation}
W(z) = \frac{1}{(1-z)\,S_\infty},
\label{eq:Wz}
\end{equation}
where $S_\infty := \lim_{z\to 1^-} S(z)$, and subsequently replace the pointwise values at $z=1$ by their finite limiting values.

To extract finite boundary coefficients, we use the \emph{regularization operator}
$\operatorname{reg}[\cdot]$,
which removes the singular part of a quantity and keeps its regular (finite) contribution at $z=1$. We therefore define
\begin{equation}
a_k^{(W)} := \operatorname{reg}[W\,a_k],
\qquad
b_k^{(W)} := \operatorname{reg}[W\,b_k],
\end{equation}
and the resulting \emph{Jansen boundary checks} at the AdS boundary become
\begin{align}
\lim_{z\to 1} a_0^{(W)} &= -1,\\
\lim_{z\to 1} a_1^{(W)} &=
\lim_{z\to 1} a_2^{(W)} =
\lim_{z\to 1} b_0^{(W)} =
\lim_{z\to 1} b_1^{(W)} = 0,
\end{align}
which enforce the normalizability condition $\Phi(1)=0$ and guarantee a discrete spectrum with $\mathrm{Im}[\Lambda] < 0$.

For the present NLED background, the large–$r$ limit of $S(r)$ gives
\begin{align}
 S_\infty^{(e)} &= 8\,\beta^{2} Q_e^{2},
 \qquad\text{(purely electric sector)},\\[2mm]
 S_\infty^{(m)} &= 8\,\beta^{2} Q_m^{2},
 \qquad\text{(purely magnetic sector)}.
\end{align}
This regularized IEF--GEVP is then discretized by Chebyshev collocation on $z\in[0,1)$~\cite{Trefethen:2000,WeidemanReddy:2000,ChebfunGuide:2014}, producing a linear matrix pencil that automatically incorporates the ingoing behavior at $r=r_+$ and the normalizability condition at the AdS boundary.

\subsection{Purely electric and purely magnetic sectors}
\label{sec:sectors}

In NLED, the axial perturbation dynamics splits naturally into two independent sectors, determined by whether the background carries electric or magnetic charge. In this subsection we outline the structural differences between these sectors for the NLEDBH under consideration, together with their implications for the effective potential and the resulting QNM spectra.

\subsubsection{Purely electric sector ($Q_m=0$)}

Specializing the dyonic lapse function to the purely electric configuration yields
\begin{equation}
f(r)
= 1 - \frac{2M}{r} + \frac{Q_e^2}{r^2}
  - 2\beta Q_e^2 + \beta^2 Q_e^2 r^2,
\label{eq:f_electric}
\end{equation}
which interpolates between the RN case ($\beta = 0$) and an effective AdS asymptotic behavior induced by the $+\beta^2 Q_e^2 r^2$ term. This family corresponds to the static limit of the Garc\'{i}a--D\'{i}az NLED black holes~\cite{Garcia-Diaz:2021,DiazGarcia:2022jpc}.\footnote{For the rotating NLED solution and its static limit, see the same references.}

For the model specified in Sec.~\ref{sec:perturbations}, we introduce the auxiliary variables
\begin{equation}
Y := \frac{L_F'}{L_F},
\qquad
Z := \frac{L_F''}{L_F},
\end{equation}
which, together with the definition $X \equiv \beta r^2$, take the explicit forms
\begin{subequations}
\begin{align}
Y &= -\frac{2\beta r\,(3+X)}{(1-X)(1+X)},\\[4pt]
Z &= -\frac{2\beta\,(3 - 15X - 3X^2 - X^3)}{(1-X)^2(1+X)^2}.
\end{align}
\end{subequations}
Substituting these expressions into Eq.~\eqref{eq:CNLED} gives
\begin{equation}
\mathcal{C}_{\rm NLED}
   = f Y^2 - 2(f'Y + fZ),
\label{eq:CNLED_1}
\end{equation}
from which the large-$r$ limit relevant for the IEF regularization follows as
\begin{equation}
S_\infty^{(e)} = 8\,\beta^2 Q_e^2,
\label{eq:Sinf-electric}
\end{equation}
namely the value used in the weight function $W(z)$ defined in Eq.~\eqref{eq:Wz}.

As expected, in the linear (Maxwell) limit, the full NLED structure collapses to the standard axial electromagnetic potential \eqref{eq:V(r)_old} on RN/Schwarzschild backgrounds; see, for instance, the AdS QNM analysis in Ref.~\cite{horowitz_hubeny_2000}.

\subsubsection{Purely magnetic sector ($Q_e=0$)}

In the purely magnetic configuration, the lapse function takes the form
\begin{equation}
f(r)
= 1 - \frac{2M}{r} + \frac{Q_m^2}{r^2}
  - 2\beta Q_m^2 + \beta^2 Q_m^2 r^2,
\label{eq:f_magnetic}
\end{equation}
which again corresponds to the static NLED background described in Refs.~\cite{Garcia-Diaz:2021,DiazGarcia:2022jpc}.  
In contrast to the electric case, the axial electromagnetic potential acquires an additional contribution proportional to $Q_m^2 L_{FF}$ that is \emph{unique} to the magnetic sector in NLED~\cite{toshmatov_electromagnetic_2018}.  
For the present model, this yields the angular factor
\begin{equation}
\frac{\ell(\ell+1)}{r^2}
\left(1 + \frac{4Q_m^2 L_{FF}}{r^4}\right)
= \frac{\ell(\ell+1)}{r^2}\,
   \frac{1 - 3\beta r^2}{1 + \beta r^2},
\label{eq:magnetic_factor}
\end{equation}
and consequently the structural function
\begin{equation}
S_m(r)
= \frac{\ell(\ell+1)}{r^2}\,
  \frac{1 - 3\beta r^2}{1 + \beta r^2}
  - \mathcal{C}_{\rm NLED}(r).
\end{equation}
The large-$r$ limit feeding the IEF asymptotic regularization gives
\begin{equation}
S_\infty^{(m)} = 8\,\beta^2 Q_m^2,
\label{eq:Sinf-magnetic}
\end{equation}
identical in structure to the purely electric sector.  
In the Maxwell limit, the factor in Eq.~\eqref{eq:magnetic_factor} reduces to unity, and the standard RN (or RN--AdS) axial electromagnetic potential is fully recovered.  
More broadly, NLED breaks the parity isospectrality between axial and polar electromagnetic perturbations and induces systematic shifts in both $\omega_R\equiv\mathrm{Re}[\omega]$ and $-\omega_I \equiv - \mathrm{Im}[\omega]$ relative to the Maxwell case; see, e.g., Refs.~\cite{toshmatov_polar_2018,Nomura:2022prd} for explicit demonstrations in NLED backgrounds.

\vspace{0.25cm}

\noindent\textbf{Remarks on well-posedness.}  
The denominators $(1 \pm \beta r^2)$ appearing in the ratios $Y$ and $Z$, motivate the working window adopted in Sec.~\ref{sec:pipeline}.  
In particular, any configuration for which $L_F$ or $L_{FF}$ vanishes or diverges outside $r_+$ must be excluded, as such cases spoil the hyperbolicity of the master equation on the physical domain $r \in [r_+, \infty)$.  
Furthermore, the effective AdS growth $f(r) \sim \beta^2 Q^2 r^2$ at large $r$ is precisely what ensures that the IEF regularization procedure introduced in Secs.~\ref{sec:ef_compactification}--\ref{sec:regularization} is applicable and numerically stable~\cite{Garcia-Diaz:2021}.

\subsection{Pseudospectral discretization and matrix GEVP}
\label{sec:spectral}

To solve the regularized eigenvalue problem, we employ a \emph{Chebyshev--Lobatto pseudospectral collocation scheme}. The compact domain $z\in[0,1]$ is discretized using the $N{+}1$ Lobatto grid points
\begin{equation}
z_j=\frac{1-\cos\!\left(\frac{\pi j}{N}\right)}{2},\qquad j=0,1,\dots,N,
\label{eq:collocation}
\end{equation}
and derivatives are approximated by the standard first- and second-order Chebyshev differentiation matrices $D$ and $D^{(2)}$ \cite{Trefethen:2000,WeidemanReddy:2000,ChebfunGuide:2014}. After implementing the asymptotic IEF regularization ($a_k\!\to\! a_k^{(W)}$, $b_k\!\to\! b_k^{(W)}$; see Subsecs.~\ref{sec:ef_compactification}--\ref{sec:regularization}), the differential operator \eqref{eq:genEVP} is represented on this grid by the matrix blocks
\begin{eqnarray}
\mathbf{A} &=& \mathrm{diag}(a_2)\,D^{(2)}
            + \mathrm{diag}(a_1)\,D
            + \mathrm{diag}(a_0), \label{eq:bfA}
\\[0.4em]
\mathbf{B} &=& \mathrm{diag}(b_1)\,D
            + \mathrm{diag}(b_0),
\label{eq:bfB}
\end{eqnarray}
obtained via pointwise evaluation of the regularized coefficients \cite{Trefethen:2000,WeidemanReddy:2000}. The Dirichlet boundary condition at $z=1$ is enforced through the rescaling \eqref{eq:Psi}, while ingoing regularity at the horizon is automatically encoded in the IEF formulation. This is the core mechanism of the linear Jansen-type approach to AdS QNMs \cite{jansen_2017}.

\smallskip
The continuous equation \eqref{eq:genEVP} then reduces to the finite-dimensional generalized eigenvalue problem
\begin{equation}
\left(\mathbf{A}+\mathrm{i}\Lambda\,\mathbf{B}\right)\vec{\Phi}=0,
\label{eq:matrixEVP}
\end{equation}
which we solve using dense linear-algebra routines. The physical frequency follows from
\begin{equation}
\omega=\Lambda\,\lvert f'(r_+)\rvert,
\label{eq:omega}
\end{equation}
and spectral convergence is monitored by increasing $N$. In practice, we employ the \texttt{QNMspectral} \textsc{Mathematica} package \cite{qnmspectral_repo,jansen_2017}, which provides an optimized implementation of the IEF--pseudospectral pipeline using linear algebra package (LAPACK)-style solvers \cite{LAPACK:1999,GolubVanLoan:2013}.

\smallskip
We briefly comment on the numerical strategy. The Chebyshev--Lobatto method guarantees exponential convergence for smooth solutions and naturally clusters points near $z=0$ and $z=1$, where the fields exhibit the steepest radial variation in these coordinates \cite{Trefethen:2000,ChebfunGuide:2014}. The differentiation matrices $D$ and $D^{(2)}$ are obtained from closed-form expressions for Chebyshev cardinal functions and are standard tools for ODE and PDE eigenvalue problems \cite{WeidemanReddy:2000}. Boundary conditions are built into the formulation: the ansatz $\psi\,e^{-i\omega v}$ ensures horizon-regular ingoing behavior, while the rescaling \eqref{eq:Psi} enforces normalizable falloff at the AdS boundary, yielding a discrete spectrum \cite{horowitz_hubeny_2000,jansen_2017}. Finally, the frequency scaling $\Lambda$ linearizes the eigenvalue dependence and significantly improves conditioning near extremality, permitting a robust extraction of QNMs even in strongly curved regimes \cite{jansen_2017}.

\subsection{Numerical pipeline and selection criteria}
\label{sec:pipeline}

We work in units where $M=1$ and focus on the dominant angular sector $\ell=2$. 
The nonlinear parameter and charge are scanned over the representative sets 
$\beta\in\{0.3,0.6,0.9\}$ and $Q\in\{0.3,0.6,0.9\}$, where $Q$ denotes $Q_e$ in the 
electric configuration and $Q_m$ in the magnetic one. The full numerical pipeline 
is consistent with the IEF compactification and regularization introduced in 
Subsecs.~\ref{sec:ef_compactification}--\ref{sec:regularization}, as well as with the 
sector decomposition described in Subsec.~\ref{sec:sectors}.

\medskip

\noindent\textbf{Parameter domain and admissibility.}  
In electromagnetic perturbations of NLED backgrounds, the coefficients of the 
axial master operator depend explicitly on $L_F$ and $L_{FF}$. In the model used 
here, the second derivative $L_{FF}$ vanishes at $r=1/\sqrt{\beta}$, which may 
introduce spurious singular structure if this radius lies in the exterior region. 
To avoid such pathologies—and to ensure that hyperbolicity, causality, and 
well-posedness remain intact on $r\in[r_+,\infty)$—we restrict attention to 
configurations satisfying $1/\sqrt{\beta}<r_+$. This guarantees that the 
zero of $L_{FF}$ is hidden behind the horizon and cannot affect the spectral 
problem. Parity non-isospectrality (axial vs polar) is a generic feature of NLED 
perturbations, as documented in 
\cite{toshmatov_electromagnetic_2018,toshmatov_polar_2018,Nomura:2022prd}, and 
fits naturally into the present analysis. For broader causality criteria in 
Plebański-class NLED, see Ref.~\cite{Schellstede:2016}.

\medskip

\noindent\textbf{Horizon determination and near-horizon scaling.}  
For every pair $(\beta,Q)$ we compute the outer horizon $r_+$ as the largest real 
root of $f(r)=0$ and evaluate the derivative $f'(r_+)$ at high precision. The 
auxiliary scaling parameter $\Lambda$ entering the IEF formulation improves the 
conditioning of the spectral problem, most notably near extremality, consistent 
with Ref.~\cite{jansen_2017}.

\medskip

\noindent\textbf{Precision scheme.}  
All computations are carried out in arbitrary precision within 
\textsc{Mathematica}, using $\texttt{prec}=60$ digits throughout the pipeline. 
The determination of $r_+$, the evaluation of $f'(r_+)$, the construction of the 
regularized coefficients, and the solution of the matrix GEVP use 
\texttt{WorkingPrecision} $=\texttt{prec}$ and 
\texttt{AccuracyGoal} $=\texttt{PrecisionGoal}=\texttt{prec}-20$, yielding 
approximately $\mathcal{O}(40)$ stable digits in intermediate quantities.  
Reported frequencies $M\omega$ in 
Tabs.~\ref{tab:qnm_electric_fund_l2}–\ref{tab:qnm_magnetic_fund_l2} are rounded to 
five decimals, well within the numerically verified convergence window; additional 
diagnostics are provided in Appendix~\ref{app:tables}.

\medskip

\noindent\textbf{Spectral construction and matrix assembly.}  
With $r_+$ fixed, the IEF-regularized coefficients $a_k^{(W)}$ and $b_k^{(W)}$ are 
evaluated on two Chebyshev–Lobatto grids to assess spectral convergence.  
We employ grid sizes $N\in\{80,100\}$ and $N\in\{120,150\}$, from which we build 
the first- and second-derivative matrices $D$ and $D^{(2)}$, together with the 
diagonal matrices $\mathrm{diag}(a_k^{(W)})$ and $\mathrm{diag}(b_k^{(W)})$.  
This follows the implementation strategy of Refs. \cite{Trefethen:2000,WeidemanReddy:2000,jansen_2017}.

\medskip

\noindent\textbf{Mode extraction and physical filtering.}  
Solving the matrix equation \eqref{eq:matrixEVP}, 
on both grids yields two approximations to the spectrum from which the physical 
frequency is reconstructed using Eq. \eqref{eq:omega}
following the IEF prescription with AdS boundary conditions 
\cite{horowitz_hubeny_2000,jansen_2017}.  
Only solutions with $\omega_I<0$ (damped) and $\omega_R\ge 0$ are retained.  
The fundamental mode is identified as the root with the smallest $|\omega_I|$, 
with ties broken by larger $\omega_R$.

\medskip

\noindent\textbf{Convergence and diagnostics.}  
To quantify spectral convergence we compare frequencies between the low- and 
high-resolution grids via
\begin{equation}
\Delta\omega=\bigl|\omega_{N_{\rm low}}-\omega_{N_{\rm high}}\bigr|,
\qquad
\texttt{digits}=-\log_{10}|\Delta\omega|.
\end{equation}
We classify results as \emph{Good} for $\texttt{digits}\ge 5$, \emph{OK} for 
$3\le\texttt{digits}<5$, and \emph{Bad} otherwise.  If the two grids return 
identical values, we set $\texttt{digits}=\infty$ but retain the qualitative 
label according to the above thresholds. The quality factor of the fundamental 
mode is quoted as
\begin{equation}
\zeta=\frac{\omega_R}{2|\omega_I|},
\end{equation}
measuring the number of oscillations per $e$-fold of decay.

\medskip

\noindent
All numerical scans are performed using \texttt{QNMspectral}~\cite{qnmspectral_repo}, 
which implements the IEF–pseudospectral framework described in 
Secs.~\ref{sec:ef_compactification}–\ref{sec:spectral}.  
The resulting fundamental ($n=0$) modes are listed in 
Tabs.~\ref{tab:qnm_electric_fund_l2} and \ref{tab:qnm_magnetic_fund_l2}, where we 
report $M\omega_R$ and $-M\omega_I$ to five decimals; full diagnostics 
($\Lambda$, $\zeta$, \texttt{digits}, $\Delta\omega$) are provided in 
Appendix~\ref{app:tables}.

\begin{figure}[t]
    \centering
    \includegraphics[width=7.0cm]{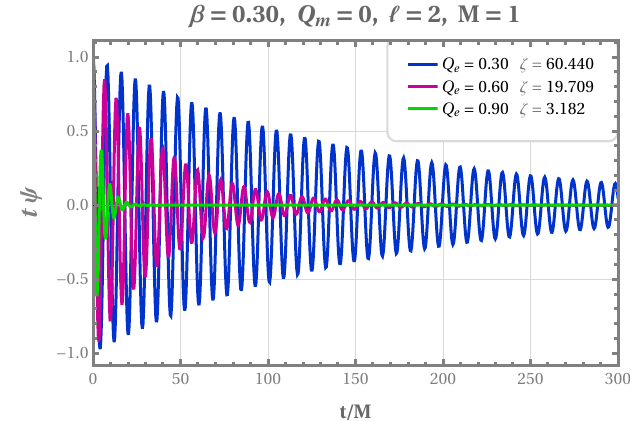} (a)\qquad
     \includegraphics[width=7.0cm]{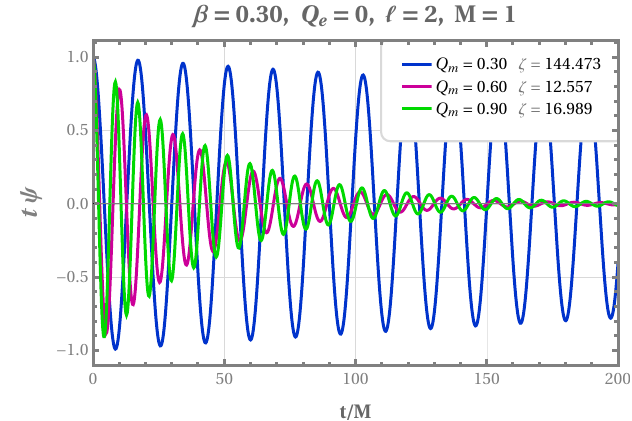} (b)
      \includegraphics[width=7.0cm]{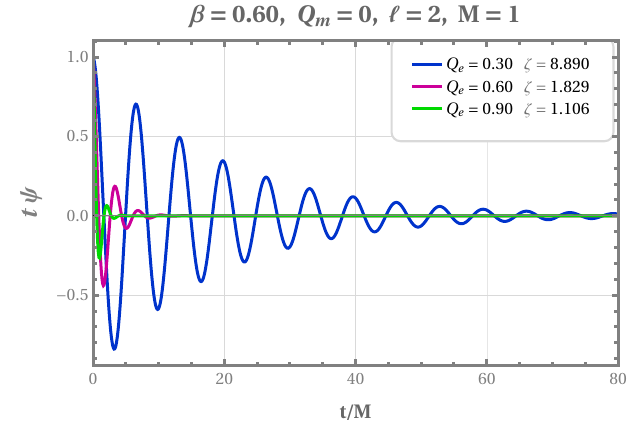} (c)\qquad
       \includegraphics[width=7.0cm]{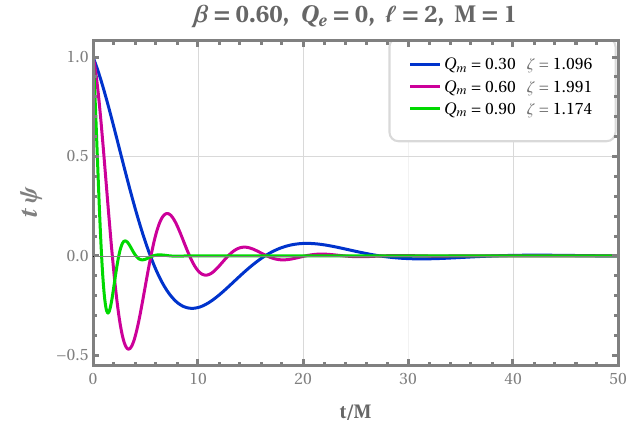} (d)
       \includegraphics[width=7.0cm]{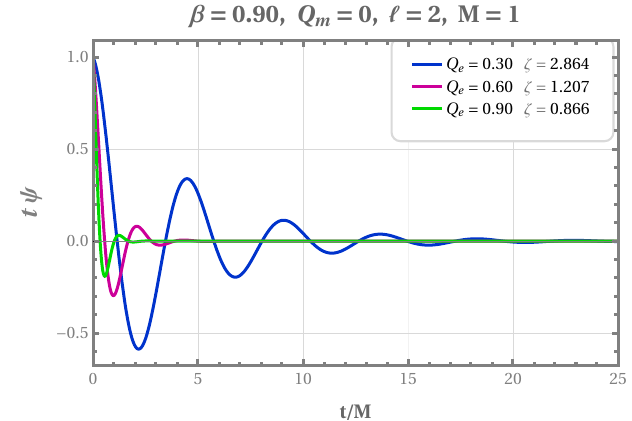} (e)\qquad
       \includegraphics[width=7.0cm]{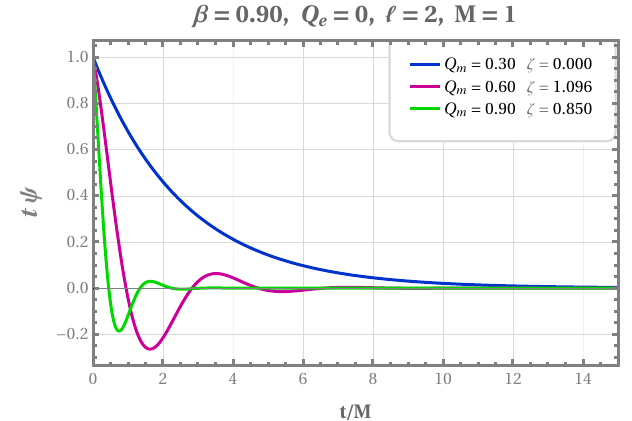} (f)
    \caption{Time-domain ringdown waveform $\psi(t)/M$ as a function of the dimensionless time $t/M$ for axial electromagnetic perturbations with $\ell=2$.  
Rows correspond to $\beta=0.30$ (a,b), $0.60$ (c,d), and $0.90$ (e,f).  
Left column: purely electric sector ($Q_m=0$); right column: purely magnetic sector ($Q_e=0$).  
Within each panel, curves are shown for $Q\in\{0.3,0.6,0.9\}$ with $M=1$.  
The waveforms are normalized single-mode damped sinusoids,
$\psi(t)=e^{-\omega_I t}\cos(\omega_R t)$, constructed from the fundamental quasinormal frequency
$\omega=\omega_R - \mathrm{i}\,\omega_I$ obtained using a Chebyshev pseudospectral discretization of the
IEF-regularized generalized eigenvalue problem.  
The legend displays the corresponding quality factor $\zeta$.  
General trends: increasing either $\beta$ or $Q$ enhances both the oscillation frequency $\omega_R$
and the damping rate $-\omega_I$.  
For fixed $(\beta,Q)$ the magnetic sector exhibits systematically smaller $\omega_R$ and $-\omega_I$ than the electric sector.  
A notable case occurs at $\beta=0.90$ and $Q_m=0.30$ (panel f), where the fundamental mode becomes purely decaying with $\omega_R\simeq 0$.}
    \label{fig:ring_down}
\end{figure}

The qualitative impact of the nonlinear coupling on the axial electromagnetic ringdown is illustrated in
Fig.~\ref{fig:ring_down}.  
For representative values of $\beta$ and charge, the time-domain signals reconstructed from the
fundamental quasinormal frequency display the characteristic behavior anticipated from the effective
potentials discussed above: increasing either $\beta$ or $Q$ yields faster oscillations and stronger
damping, while the magnetic sector systematically shows lower $\omega_R$ and $-\omega_I$ than the
electric one for identical parameter choices.  
In particular, the overdamped case $(\beta,Q_m)=(0.90,0.30)$, where the fundamental mode becomes
purely imaginary, manifests in the waveform as a purely exponential decay with no oscillatory regime
(cf.\ panel~f).

%
\begin{table}[h!]
  \centering
  \caption{Fundamental quasinormal mode ($n=0$) in the electric sector ($Q_m=0$).
  Listed are the dimensionless frequencies $M\omega$ for fixed $\ell=2$ and $M=1$.
  Each row specifies the nonlinearity parameter $\beta$, the electric charge $Q_e$,
  and the corresponding real and imaginary parts of the frequency.}
  \label{tab:qnm_electric_fund_l2}

  \begin{tabular}{
    S[table-format=1.1]
    @{\hspace{1.2em}}
    S[table-format=1.1]
    @{\hspace{0.8em}}
    S[table-format=1.5]
    S[table-format=1.5]
  }
    \toprule
    {$\beta$} & {$Q_e$} & {$M\omega_R$} & {$-M\omega_I$} \\
    \midrule
    0.3 & 0.3 & 0.77889 & 0.00644 \\
    0.3 & 0.6 & 0.94659 & 0.02401 \\
    0.3 & 0.9 & 1.29777 & 0.20392 \\
    0.6 & 0.3 & 0.95041 & 0.05346 \\
    0.6 & 0.6 & 1.79832 & 0.49152 \\
    0.6 & 0.9 & 2.81319 & 1.27132 \\
    0.9 & 0.3 & 1.36566 & 0.23841 \\
    0.9 & 0.6 & 2.82767 & 1.17094 \\
    0.9 & 0.9 & 4.64806 & 2.68445 \\
    \bottomrule
  \end{tabular}

  \vspace{3pt}
  \raggedright\footnotesize\textit{Notes:}  
  We list separately the real part $M\omega_R$ and the magnitude of the imaginary part $-M\omega_I>0$, corresponding to damped modes. 
  Reported values are rounded to five significant decimal places, consistent with the verified cross-grid convergence for the fundamental mode.
\end{table}
%
\begin{table}[h!]
  \centering
  \caption{Fundamental quasinormal mode ($n=0$) in the magnetic sector ($Q_e=0$).
  Frequencies $M\omega$ are given in dimensionless form for fixed $\ell=2$ and $M=1$.
  Columns list the nonlinearity parameter $\beta$, the magnetic charge $Q_m$, and the real and imaginary components of the frequency.}
  \label{tab:qnm_magnetic_fund_l2}

  \begin{tabular}{
    S[table-format=1.1]
    @{\hspace{1.2em}}
    S[table-format=1.1]
    @{\hspace{0.8em}}
    S[table-format=1.5]
    S[table-format=1.5]
  }
    \toprule
    {$\beta$} & {$Q_m$} & {$M\omega_R$} & {$-M\omega_I$} \\
    \midrule
    0.3 & 0.3 & 0.36638 & 0.00127 \\
    0.3 & 0.6 & 0.61700 & 0.02457 \\
    0.3 & 0.9 & 0.73556 & 0.02165 \\
    0.6 & 0.3 & 0.28756 & 0.13124 \\
    0.6 & 0.6 & 0.85411 & 0.21452 \\
    0.6 & 0.9 & 1.91955 & 0.81724 \\
    0.9* & 0.3 & 0.00000 & 0.38833 \\
    0.9 & 0.6 & 1.66480 & 0.75937 \\
    0.9 & 0.9 & 3.50265 & 2.05927 \\
    \bottomrule
  \end{tabular}

  \vspace{3pt}
  \raggedright\footnotesize\textit{Notes:} An asterisk ($^{*}$) denotes a \emph{purely imaginary} fundamental mode, i.e.\ $\omega_R=0$
  within numerical precision (here at $\beta=0.9$, $Q_m=0.3$), corresponding to an overdamped ringdown.
  This parameter choice coincides with the regime in which the magnetic effective potential becomes
  \emph{monotonic}, exhibiting no near-horizon barrier or trapping region (see Fig.~\ref{fig:Vem}).  
  This is consistent with the general expectation that the oscillatory part of the QNM frequency in the eikonal limit is governed by the height of a potential barrier (or, equivalently, by the properties of the unstable photon orbit). 
  Configurations lacking such a barrier may develop non-oscillatory, purely imaginary modes; see, e.g., Refs. \cite{Cardoso_2009PRD,Konoplya_2023PLB,Berti_2009CQG}.
  Reported values are rounded to five decimals, consistent with the verified cross-grid accuracy for the fundamental mode.
\end{table}
%
%
%

\subsection{Discussion}
\label{sec:discussion}


The inclusion of NLED introduces substantial modifications to the
quasinormal spectrum of the black hole when compared with the linear Maxwell limit.  
A systematic analysis of the fundamental mode reveals that increasing either the nonlinearity
parameter $\beta$ or the charge magnitude $Q$ leads to a higher oscillation frequency and a markedly
enhanced damping rate (i.e., larger $-\omega_I$).  
Consequently, the ringdown becomes characterized by shorter oscillation periods and more rapid decay
as nonlinear electromagnetic effects strengthen.

A second key result is the pronounced distinction between the spectra associated with purely electric
and purely magnetic backgrounds.  In contrast to linear electrodynamics—where isospectrality between
these configurations would hold—the nonlinear theory breaks this degeneracy.  
For identical values of $(M,Q)$, modes in the magnetic sector generally exhibit smaller
$\omega_R$ and reduced damping compared to their electric-sector counterparts.
Moreover, for sufficiently large $\beta$ and comparatively small $Q_m$, the fundamental mode becomes
purely imaginary, with $\omega_R \approx 0$ within numerical precision.  
This corresponds to a purely decaying, overdamped perturbation.  
The emergence of such non-oscillatory behavior correlates directly with the disappearance of the
near-horizon potential barrier, i.e., with the loss of an unstable photon ring, precisely as expected
from general principles of black hole perturbation theory
\cite{Cardoso_2009PRD,Konoplya_2023PLB,Berti_2009CQG}.  
The appearance of these overdamped modes therefore represents a genuinely new physical feature of
nonlinearly charged black holes and may serve as a characteristic signature of NLED effects.

Taken together, our results are consistent with and complementary to earlier studies showing that
nonlinear electromagnetic corrections break the parity-related spectral degeneracy
\cite{Nomura:2022prd,toshmatov_electromagnetic_2018,toshmatov_polar_2018}, and that increasing
nonlinearity tends to enhance the damping of quasinormal oscillations
\cite{Nomura:2022prd}.  
At the same time, we uncover qualitatively new behavior—most notably, the onset of purely
imaginary fundamental modes in the magnetic sector—that goes beyond simple shifts in the spectrum.  
These findings highlight the importance of incorporating NLED effects into black hole perturbation
studies, both for accurate theoretical modeling and for potential observational diagnostics in
highly magnetized or strongly charged astrophysical environments.

\section{Summary and conclusions}
\label{sec:conclusions}


In this work we analyzed the axial electromagnetic QNMs of a static black hole in a Pleba\'{n}ski-class NLED
theory. By combining an IEF reformulation with a
compactified radial coordinate and an asymptotic regularization of the coefficients, we cast the perturbation equation into a linear generalized eigenvalue problem solvable through Chebyshev--Lobatto pseudospectral collocation. This framework provides stable and exponentially convergent spectra across the full parameter domain compatible with a well-behaved NLED
background.

Our results demonstrate that the NLED sector introduces
significant modifications to the QNM spectrum. Increasing the
nonlinearity parameter $\beta$ or the charge magnitude $Q$ enhances both the oscillation frequency and the damping rate of the fundamental mode, leading to shorter-lived and more rapidly decaying ringdowns compared to the Maxwell limit. We also find a clear spectral distinction between purely electric and
purely magnetic backgrounds, reflecting the breakdown of parity isospectrality well known in NLED theories. Modes in the magnetic sector are generally less oscillatory and more weakly damped for the same $(\beta,Q)$.

A striking feature arises for large $\beta$ and moderately small $Q_m$, where the effective potential becomes monotonic and the fundamental QNM turns purely imaginary. This overdamped, non-oscillatory decay correlates with the absence of a photon sphere barrier, in agreement with general arguments relating the
eikonal QNM frequency to unstable null geodesics. Such behavior is absent in linear electrodynamics and may serve as a distinctive signature of strong nonlinear effects.

To conclude, our analysis shows that NLED can induce both
quantitative and qualitative modifications in black hole ringdown. These results motivate further studies including polar perturbations, full parity-breaking spectra, and rotating or dynamically evolving NLED black holes, as well as possible observational consequences for compact objects immersed in
strong electromagnetic fields.

\begin{acknowledgments}
M.F. is supported by Universidad Central de Chile through project No. PDUCEN20240008.  J.R.V. is partially supported by Centro de F\'isica Teórica de Valparaíso (CeFiTeV). 
\end{acknowledgments}

\appendix

\section{Analytical method for determining the roots of quartic equations}
\label{app:A}

Consider a general quartic equation of the form
\begin{equation}
    \mathrm{a}_4 r^4 + \mathrm{a}_3 r^3 + \mathrm{a}_2 r^2 + \mathrm{a}_1 r + \mathrm{a}_0 = 0.
    \label{eq:A0}
\end{equation}
Introducing the variable transformation $r \doteq x - \mathrm{a}_3/4$ eliminates the cubic term, leading to the depressed quartic
\begin{equation}
    x^4 + \mathcal{a} x^2 + \mathcal{b} x + \mathcal{c} = 0,
    \label{eq:A1}
\end{equation}
where the new coefficients are given by
\begin{subequations}
    \begin{align}
        \mathcal{a} &= \mathrm{a}_2 - \frac{3\mathrm{a}_3^2}{8},\\
        \mathcal{b} &= \mathrm{a}_1 + \frac{\mathrm{a}_3^3}{8} - \frac{\mathrm{a}_3 \mathrm{a}_2}{2},\\
        \mathcal{c} &= \mathrm{a}_0 + \frac{\mathrm{a}_3^2 \mathrm{a}_2}{16} - \frac{3\mathrm{a}_3^4}{256} - \frac{\mathrm{a}_3 \mathrm{a}_1}{4}.
    \end{align}
    \label{eq:A2}
\end{subequations}
The quartic equation \eqref{eq:A1} can be factorized into the product of two quadratic polynomials as
\begin{equation}
    x^4 + \mathcal{a} x^2 + \mathcal{b} x + \mathcal{c} = 
    \bigl(x^2 - 2\tilde{\alpha}x + \tilde{\beta}\bigr)
    \bigl(x^2 + 2\tilde{\alpha}x + \tilde{\gamma}\bigr),
    \label{eq:A3}
\end{equation}
which yields the following relations:
\begin{subequations}
\begin{align}
    \mathcal{a} &= \tilde{\beta} + \tilde{\gamma} - 4\tilde{\alpha}^2,\\
    \mathcal{b} &= 2\tilde{\alpha}(\tilde{\beta} - \tilde{\gamma}),\\
    \mathcal{c} &= \tilde{\beta} \tilde{\gamma}.
    \label{eq:app_Abeta}
\end{align}
\label{eq:A4}
\end{subequations}
By solving the first two equations for $\tilde{\beta}$ and $\tilde{\gamma}$, one obtains
\begin{subequations}
\begin{align}
    \tilde{\beta} &= 2\tilde{\alpha}^2 + \frac{\mathcal{a}}{2} + \frac{\mathcal{b}}{4\tilde{\alpha}},\\
    \tilde{\gamma} &= 2\tilde{\alpha}^2 + \frac{\mathcal{a}}{2} - \frac{\mathcal{b}}{4\tilde{\alpha}}.
\end{align}
\label{eq:A5}
\end{subequations}
Substituting these expressions into Eq.~\eqref{eq:app_Abeta} leads to the sextic equation for $\tilde{\alpha}$:
\begin{equation}
    \tilde{\alpha}^6 + \frac{\mathcal{a}}{2}\tilde{\alpha}^4
    + \left( \frac{\mathcal{a}^2}{16} - \frac{\mathcal{c}}{4} \right) \tilde{\alpha}^2
    - \frac{\mathcal{b}^2}{64} = 0.
    \label{eq:A6}
\end{equation}
To simplify this, we define a new variable
\begin{equation}
    \tilde{\alpha}^2 = \tilde{U} - \frac{\mathcal{a}}{6},
    \label{eq:A7}
\end{equation}
which transforms Eq.~\eqref{eq:A6} into a depressed cubic equation,
\begin{equation}
    \tilde{U}^3 - \tilde{\eta}_2\,\tilde{U} - \tilde{\eta}_3 = 0,
    \label{eq:A8}
\end{equation}
where
\begin{subequations}
\begin{align}
    \tilde{\eta}_2 &= \frac{\mathcal{a}^2}{48} + \frac{\mathcal{c}}{4},\\
    \tilde{\eta}_3 &= \frac{\mathcal{a}^3}{864} + \frac{\mathcal{b}^2}{64} - \frac{\mathcal{a}\mathcal{c}}{24}.
\end{align}
\label{eq:A9}
\end{subequations}
The real root of this cubic equation can be expressed as \cite{Nickalls:2006,Zucker:2008}
\begin{equation}
    \tilde{U} = 2\sqrt{\frac{\tilde{\eta}_2}{3}}
    \cosh\!\left[
        \frac{1}{3} \arccosh\!\left(
        \frac{3\tilde{\eta}_3}{2}\sqrt{\frac{3}{\tilde{\eta}_2^3}}
        \right)
    \right].
    \label{eq:A10}
\end{equation}
Finally, substituting $\tilde{\alpha}$, $\tilde{\beta}$, and $\tilde{\gamma}$ back into the factorized form, the four roots of Eq.~\eqref{eq:A1} are given by
\begin{eqnarray}
    x_1 &=& \tilde{\alpha} + \sqrt{\tilde{\alpha}^2 - \tilde{\beta}}, \label{eq:A11}\\
    x_2 &=& \tilde{\alpha} - \sqrt{\tilde{\alpha}^2 - \tilde{\beta}}, \label{eq:A12}\\
    x_3 &=& -\tilde{\alpha} + \sqrt{\tilde{\alpha}^2 - \tilde{\gamma}}, \label{eq:A13}\\
    x_4 &=& -\tilde{\alpha} - \sqrt{\tilde{\alpha}^2 - \tilde{\gamma}}. \label{eq:A14}
\end{eqnarray}
Hence, the corresponding roots of the original quartic equation \eqref{eq:A0} can be expressed as
\begin{equation}
    r_j = x_j - \frac{\mathrm{a}_3}{4}, 
    \qquad j = \overline{1,4}.
\end{equation}

\section{Full quasinormal-mode tables}
\label{app:tables}

These tables complement Tabs.~\ref{tab:qnm_electric_fund_l2} and
\ref{tab:qnm_magnetic_fund_l2} by reporting the rescaled eigenvalues
$\Lambda$, the quality factors $\zeta$, and the convergence diagnostics (\texttt{digits}, $\Delta\omega$) associated with all fundamental modes. 

\begin{table*}[htbp]
\centering
\caption{Electric sector ($Q_m=0$): full QNM data for the fundamental mode ($n=0$, $M=1$, $\ell=2$). For each configuration we report the rescaled eigenvalue $\Lambda$, the dimensionless frequency $M\omega$, the quality factor $\zeta$, and the convergence diagnostics \texttt{digits} and $\Delta\omega$, obtained from the cross-resolution comparison described in Subsec.~\ref{sec:pipeline}.}
\label{tab:qnm_electric_full}
\begin{tabular}{
  S[table-format=1.2] 
  S[table-format=1.2] 
  S[table-format=1.6] 
  S[table-format=1.6] 
  S[table-format=1.5] 
  S[table-format=1.5] 
  S[table-format=2.6] 
  S[table-format=2.6] 
  c                  
  c                  
}
\toprule
 {$\beta$} & {$Q_e$} &
 {$\mathrm{Re}(\Lambda)$} & {$-\mathrm{Im}(\Lambda)$} &
 {$M\omega_R$} & {$-M\omega_I$} &
 {$\zeta$} & {digits} & {quality} & {$\Delta\omega$} \\
\midrule
0.3 & 0.3 &
1.52517 & 0.01262 &
0.77889 & 0.00644 &
60.44030 & 8.77821 & \texttt{Good} & $1.66645\times10^{-9}$ \\[2pt]
0.3 & 0.6 &
1.74774 & 0.04434 &
0.94659 & 0.02401 &
19.70870 & 10.07720 & \texttt{Good} & $8.37233\times10^{-11}$ \\[2pt]
0.3 & 0.9 &
2.20033 & 0.34574 &
1.29777 & 0.20392 &
3.18209 & 8.19775 & \texttt{Good} & $6.34239\times10^{-9}$ \\[2pt]
0.6 & 0.3 &
1.49116 & 0.08387 &
0.95041 & 0.05346 &
8.88952 & 9.49992 & \texttt{Good} & $3.16288\times10^{-10}$ \\[2pt]
0.6 & 0.6 &
1.87784 & 0.51325 &
1.79832 & 0.49152 &
1.82935 & 6.89188 & \texttt{Good} & $1.28268\times10^{-7}$ \\[2pt]
0.6 & 0.9 &
2.06800 & 0.93456 &
2.81319 & 1.27132 &
1.10641 & 4.76042 & \texttt{OK}   & $1.73610\times10^{-5}$ \\[2pt]
0.9 & 0.3 &
1.61877 & 0.28260 &
1.36566 & 0.23841 &
2.86409 & 8.07621 & \texttt{Good} & $8.39054\times10^{-9}$ \\[2pt]
0.9 & 0.6 &
1.83834 & 0.76126 &
2.82767 & 1.17094 &
1.20744 & 5.40807 & \texttt{Good} & $3.90779\times10^{-6}$ \\[2pt]
0.9 & 0.9 &
1.97241 & 1.13915 &
4.64806 & 2.68445 &
0.86574 & 3.52133 & \texttt{OK}   & $3.01074\times10^{-4}$ \\
\bottomrule
\end{tabular}
\end{table*}
\vspace{3pt}
\begin{table*}[htbp]
\centering
\caption{Magnetic sector ($Q_e=0$): full QNM data for the fundamental mode ($n=0$, $M=1$, $\ell=2$). Column definitions are identical to those in Tab.~\ref{tab:qnm_electric_full}, including the rescaled eigenvalue $\Lambda$, the dimensionless frequency $M\omega$, the quality factor $\zeta$, and the convergence diagnostics \texttt{digits} and $\Delta\omega$.}
\label{tab:qnm_magnetic_full}
\begin{tabular}{
  S[table-format=1.2] 
  S[table-format=1.2] 
  S[table-format=1.6] 
  S[table-format=1.6] 
  S[table-format=1.5] 
  S[table-format=1.5] 
  S[table-format=3.6] 
  S[table-format=2.6] 
  c                  
  c                  
}
\toprule
 {$\beta$} & {$Q_m$} &
 {$\mathrm{Re}(\Lambda)$} & {$-\mathrm{Im}(\Lambda)$} &
 {$M\omega_R$} & {$-M\omega_I$} &
 {$\zeta$} & {digits} & {quality} & {$\Delta\omega$} \\
\midrule
0.3 & 0.3 &
0.71741 & 0.00248 &
0.36638 & 0.00127 &
144.47300 & 10.06580 & \texttt{Good} & $8.59323\times10^{-11}$ \\[2pt]
0.3 & 0.6 &
1.13919 & 0.04536 &
0.61700 & 0.02457 &
12.55700 & 9.98675 & \texttt{Good} & $1.03099\times10^{-10}$ \\[2pt]
0.3 & 0.9 &
1.24712 & 0.03670 &
0.73556 & 0.02165 &
16.98950 & 10.82900 & \texttt{Good} & $1.48263\times10^{-11}$ \\[2pt]
0.6 & 0.3 &
0.45118 & 0.20591 &
0.28756 & 0.13124 &
1.09557 & 8.85534 & \texttt{Good} & $1.39528\times10^{-9}$ \\[2pt]
0.6 & 0.6 &
0.89188 & 0.22400 &
0.85411 & 0.21452 &
1.99077 & 9.44683 & \texttt{Good} & $3.57411\times10^{-10}$ \\[2pt]
0.6 & 0.9 &
1.41108 & 0.60076 &
1.91955 & 0.81724 &
1.17441 & 6.89616 & \texttt{Good} & $1.27011\times10^{-7}$ \\[2pt]
0.9 & 0.3 &
0.00000 & 0.46030 &
0.00000 & 0.38833 &
0.00000 & 8.88456 & \texttt{Good} & $1.30450\times10^{-9}$ \\[2pt]
0.9 & 0.6 &
1.08233 & 0.49368 &
1.66480 & 0.75937 &
1.09618 & 7.46105 & \texttt{Good} & $3.45896\times10^{-8}$ \\[2pt]
0.9 & 0.9 &
1.48636 & 0.87386 &
3.50265 & 2.05927 &
0.85046 & 5.08256 & \texttt{Good} & $8.26879\times10^{-6}$ \\
\bottomrule
\end{tabular}
\end{table*}
%

\bibliographystyle{ieeetr}
\bibliography{biblio_v1}

\end{document}